\documentclass[10pt,journal,compsoc]{IEEEtran}
\ifCLASSOPTIONcompsoc
  \usepackage[nocompress]{cite}
\else
  \usepackage{cite}
\fi
\ifCLASSINFOpdf
\else
\fi

\usepackage{amsmath,amssymb,amsfonts}
\usepackage{graphicx}
\usepackage{textcomp}
\usepackage{xcolor}
\usepackage{times}
\usepackage{xcolor}
\usepackage{soul}
\usepackage{multirow}
\usepackage{subfigure}
\usepackage[utf8]{inputenc}
\usepackage{caption}

\usepackage{diagbox}
\usepackage{bm}
\usepackage{url}

\usepackage{algorithm}

\usepackage{algpseudocode}

\begin{document}

\title{FedHome: Cloud-Edge based Personalized Federated Learning for In-Home Health Monitoring}
%
%
%
%
\author{
	Qiong Wu, Xu Chen,~\IEEEmembership{Senior Member,~IEEE}, Zhi Zhou,~\IEEEmembership{Member,~IEEE}, and Junshan Zhang,~\IEEEmembership{Fellow,~IEEE}
	\thanks{Q. Wu, X. Chen and Z. Zhou are with School of Computer Science and Engineering, Sun Yat-sen University, Guangzhou 510006, China. J. Zhang is with the School of Electrical, Computer and Energy Engineering, Arizona State University, Tempe, AZ 85287 USA.}}

\IEEEtitleabstractindextext{%
\begin{abstract}
In-home health monitoring has attracted great attention for the ageing population worldwide. With the abundant user health data accessed by Internet of Things (IoT) devices and recent development in machine learning, smart healthcare has seen many successful stories. However, existing approaches for in-home health monitoring do not pay sufficient attention to user data privacy and thus are far from being ready for large-scale practical deployment. In this paper, we propose FedHome, a novel cloud-edge based federated learning framework for in-home health monitoring, which learns a shared global model in the cloud from multiple homes at the network edges and achieves data privacy protection by keeping user data locally. To cope with the imbalanced and non-IID distribution inherent in user's monitoring data, we design a generative convolutional autoencoder (GCAE), which aims to achieve accurate and personalized health monitoring by refining the model with a generated class-balanced dataset from user's personal data. Besides, GCAE is lightweight to transfer between the cloud and edges, which is useful to reduce the communication cost of federated learning in FedHome. Extensive experiments based on realistic human activity recognition data traces corroborate that FedHome significantly outperforms existing widely-adopted methods.
\end{abstract}

\begin{IEEEkeywords}
Federated learning, in-home health monitoring, personalization.
\end{IEEEkeywords}}

\maketitle

\IEEEdisplaynontitleabstractindextext

%
\IEEEpeerreviewmaketitle

\IEEEraisesectionheading{\section{Introduction}\label{sec:introduction}}
\IEEEPARstart{W}{ith} lower fertility and longer life expectancy, population ageing is becoming a global issue. According to the World Health Organization (WHO), the world's population aged 60 years and above will increase to 1.2 billion in 2025 and subsequently to 2 billion by 2050  \cite{WHO}. The ageing population tends to have a higher prevalence of chronic diseases, physical disabilities, mental illnesses and other co-morbidities, deriving problems such as shortage of medical resources and reduction of quality in healthcare services. Moreover, some older adults (e.g., solitary elderly, elderly couples) prefer to live independently in their own homes, bringing an increased risk of falls and strokes which could be life-threatening. Therefore, there is a growing demand for developing technologies that can aid in the care of the elderly from hospital-centric to home-centric. With the proliferation of smart devices, mobile networks and computing infrastructures, Internet of Things (IoT) is poised to make substantial advances in healthcare systems due to the integrated sensing, computation and communication capabilities of IoT devices. Thus, IoT-based in-home health monitoring is envisioned as a promising paradigm and has attracted great attention \cite{mano2016exploiting, verma2018fog}.

\begin{figure}[!t]
	\centering
	\includegraphics[width=0.94\linewidth]{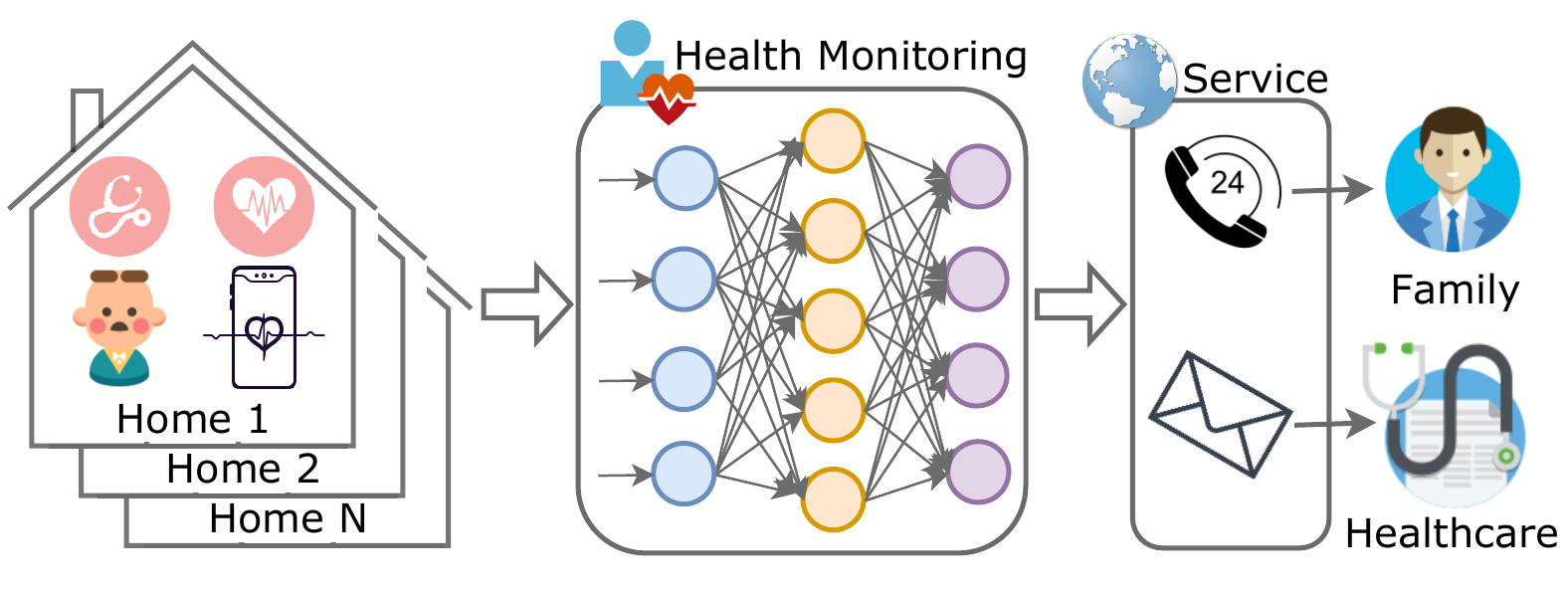}
	\caption{In-home health monitoring. A large quantity of user data from smart healthcare IoT devices in multiple homes is collected to train machine learning (ML) models for in-home health monitoring and supportive healthcare services. As the user's health data is privacy-sensitive, ML models can not be traditionally managed in a centralized manner.}
	\label{flHome}
	\vspace{-10pt}
\end{figure}

IoT devices, particularly wearable devices as exemplified by smartphones and smartwatches, can track a user's activity, heart rate and so forth persistently, and then feedback these data to a healthcare center for further processing and diagnosis.  Due to the easiness and low cost of people's health information acquisition, smart in-home healthcare is able to facilitate pervasive and nonintrusive health monitoring, and is emerging as mainstream for smart healthcare \cite{hiremath2014wearable}. Current healthcare applications often utilize machine learning (ML) models trained on massive user data to yield insights, perform in-home health monitoring and ultimately provide better public health services and products. As illustrated in Fig. \ref{flHome}, one possible way is to collect a large quantity of training data required for a powerful in-home health monitoring system from many families. As users' health data involves individual privacy, collecting and storing abundant health monitoring data at a remote cloud center to support data analytics based smart healthcare services would impose great privacy leakage risk, making people reluctant to adopt such services. At the same time, with the increasing concerns on data security and user privacy, several regulatory policies and protection mechanisms have been set to restrict data access and protect data privacy. For example, General Data Protection Regulation (GDPR), issued by the European Union, has enforced strict rules on data security and privacy protection \cite{voigt2017eu}. Additionally, China and the United States have also strengthened their attention to data privacy \cite{inkster2018china}. Under the increasingly stringent data privacy protection legislation, it is almost impossible to integrate sufficient user health data scattered around multiple families for ML model training and in-home health monitoring.

Emerging wisdom to address the above privacy challenge in the traditional, centralized machine learning approaches is federated learning (FL) proposed by Google in 2016. FL enables jointly training a shared global machine learning model under the coordination of a central server by aggregating locally-computed updates while keeping all the sensitive data in local clients (e.g., smartphones), thus allowing users to collectively reap the benefits of the shared model without compromising their data privacy. This form of learning is ideal for community healthcare because it ensures privacy by default, respects data ownership, and maintains locality of data for application deployment at scale. Furthermore, federated learning makes all the participating clients (e.g., medical institutions, smart healthcare IoT devices) share their experiences with privacy guarantee, resulting in a significantly improved performance of the ML model \cite{xu2020federated}.

Nevertheless, existing studies of FL focus on training a single global model, which would suffer from the weaknesses in both statistical and communication perspectives and lacks personalization, resulting in a degraded performance in the healthcare scenarios. Firstly, since different users have different physical characteristics, the personally-generated data naturally exhibits the kind of non-IID (a.k.a. non-independent and identically distributed) distribution. Even for a single user, the health monitoring  data can be highly skewed. For example, an adult's activity data may include a lot of standing and walking samples, but very few falls. Both the imbalanced and non-IID distribution of user health data may greatly degrade the learning performance. Secondly, the communication challenge refers to communication constraints, such as slow or expensive connections and limited bandwidth costs. Existing federated learning approaches mainly focus on the communication rounds reduction considering the fact that mobile devices are frequently offline \cite{lim2020federated}. However, the communication overload reduction in each round is also critical to improve the efficiency of the algorithm considering the bandwidth cost. Finally, as for personalization, the shared model trained by federated learning only captures the common features of all users, but it may perform poorly on a particular user. Thus, it is necessary to learn the fine-grained information on a particular user for personalized healthcare. However, current federated learning researches in healthcare only focus on one of these challenges, without considering them as an interrelated integration. For example, Zhao et al. propose a data-sharing strategy to improve the FedAvg algorithm with non-IID by creating a small globally-shared dataset \cite{zhao2018federated}. Heong et al. advocate federated distillation to reduce the communication size to transfer in FL \cite{jeong2018communication}. Chen et al. propose a federated transfer learning framework for personalized healthcare \cite{chen2020fedhealth}.

To cope with above issues, we propose FedHome, a cloud-edge federated learning framework for personalized in-home health monitoring, which is a novel scheme that can simultaneously address the three challenges in a holistic manner. FedHome utilizes a synergistic cloud-edge computing architecture and federated learning to avoid uploading users' sensitive data to a centralized cloud server, thus reducing the network latency, and more importantly, addressing privacy concerns. We devise a novel generative convolutional autoencoder (GCAE) network as the model trained on both the cloud and the edges. By integrating the parameter sharing mechanism in convolutional neural network (CNN) and representation ability of autoencoder (AE), GCAE is able to capture the representative and low-dimensional features of user data. Moreover, the representation ability of GCAE is improved together with the prediction ability through the model training process in federated settings. After global model training, GCAE generates new data samples of minority classes in low-dimension space with synthetic minority over-sampling technique (SMOTE) in order to deal with the imbalanced and non-IID distribution exhibited in users' health data. With the generated class-balanced dataset, the trained GCAE model can be personalized by fine-tuning its parameters for precise healthcare performance during classification or inference. It is worthnoting that GCAE plays a significant role in both training and personalization stages, thus resulting in an increased classification accuracy. Furthermore, GCAE is able to reduce communication overhead without compromising performance due to parameter sharing and autoencoder mechanism.

\begin{figure*}[!t]
	\centering
	\includegraphics[width=0.95\linewidth]{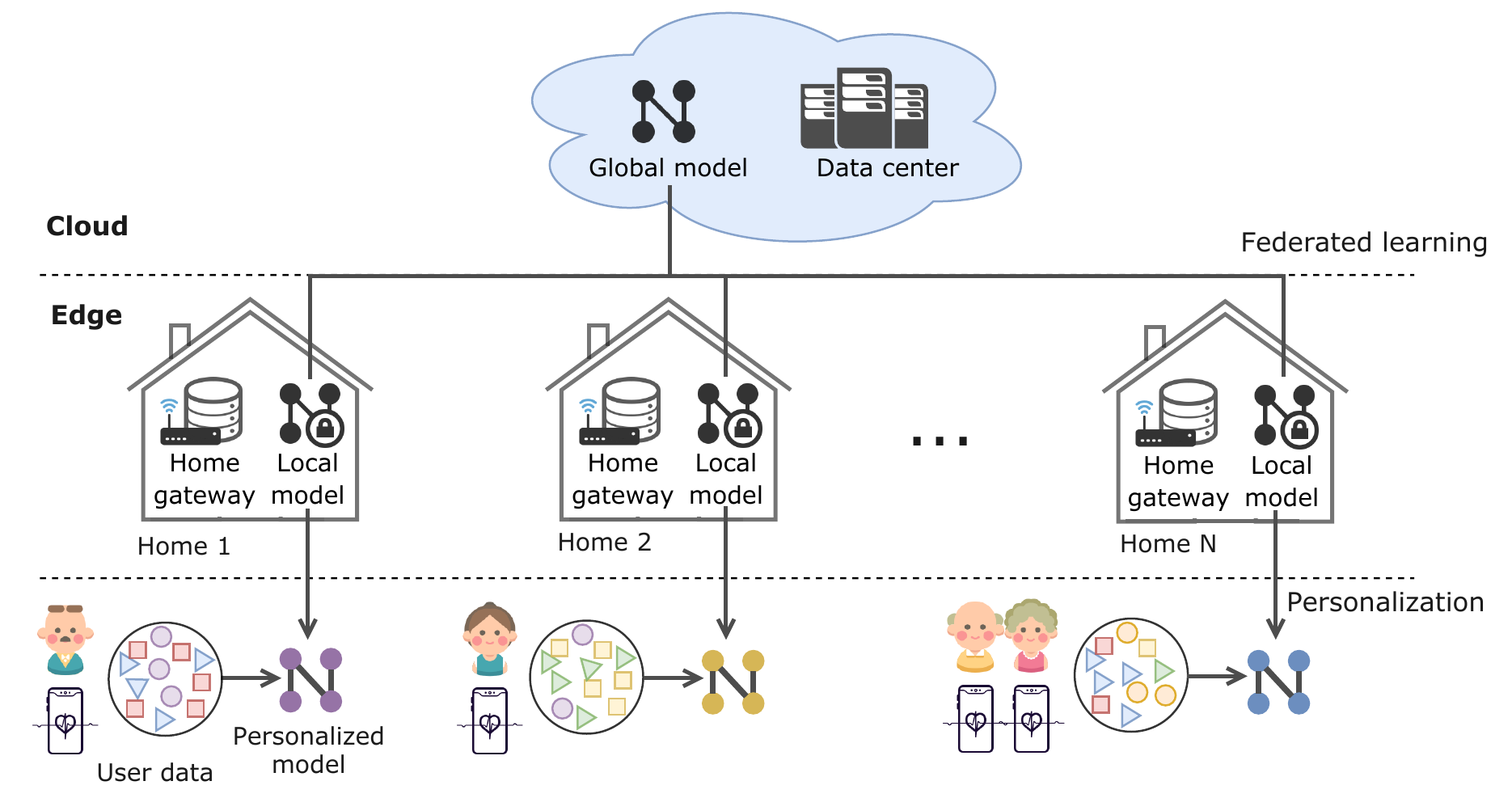}
	\vspace{-10pt}
	\caption{FedHome framework overview.}
	\label{framework}
	\vspace{-10pt}
\end{figure*}

In summary, this paper makes the following contributions:
\begin{itemize}		
	\item We propose FedHome, a novel cloud-edge based personalized federated learning framework for in-home health monitoring. FedHome trains a globally shared model from dispersed homes under the coordination of a central server in a cloud-edge paradigm, which can prevent data leakage by keeping user data locally and then achieve personalized model learning with user's local data.

	\item FedHome features on a novel design of the generative convolutional autoencoder (GCAE). By synthesizing samples of minority classes and retraining user's local model with the generated class-balanced dataset, GCAE mitigates the prediction performance degradation caused by imbalanced and non-IID distribution of user data, and achieves superior personalized predictions. Besides, GCAE is a lightweight model by parameter sharing and dimension reduction, which makes the model transfer between the cloud and the edges much more cost-efficient.
	
	\item Extensive experiments are conducted using a realistic human activity dataset to demonstrate the effectiveness of FedHome, which shows that FedHome well outperforms traditional centralized methods in both the balanced and the imbalanced data cases. For example, FedHome achieves a high accuracy of $95.41\%$ with more than $7.49\%$ accuracy improvement over the widely-used convolutional neural network (CNN) algorithm. Comparing with other federated learning approaches, FedHome achieves a gain of more than $10\%$ accuracy.
	
\end{itemize}

The rest of this paper is organized as follows: Section \ref{SectionProblemDefinition} outlines the framework overview of FedHome. In Section \ref{SectionLearning}, we introduce the algorithm details of FedHome. In Section \ref{SectionExper}, we conduct extensive experiments using a realistic human activity dataset. In Section \ref{RelatedWork}, we review the related work in in-home health monitoring and federated learning. Finally, we conclude our paper in Section \ref{SectionConclusion}.

\section{Overview of FedHome}
\label{SectionProblemDefinition}
We consider an in-home health monitoring system consisting of a cloud server and $N$ dispersed edge computing nodes (e.g., smarthome gateway) deployed at users' homes. Given a specific task, for example, elderly fall-down detection via human activity recognition, we wish to train an accurate machine learning model by taking advantage of the abundant data $\{\mathcal{D}_{1}, \mathcal{D}_{2}, \cdots, \mathcal{D}_{N}\}$ collected from all the home users in $N$ edge nodes. Conventional approaches usually put all the data together (e.g., uploading to a cloud server) to train a model $\mathcal{M}_{ALL}$ in a centralized fashion. However, such approaches would cause significant privacy issues as the health data usually contains a user's sensitive private and personal information. Thus, we resort to the federated learning paradigm, in which users collaboratively train a model $\mathcal{M}_{FED}$ without uploading their personal data to a central cloud server or exposing to each other.

There have been some attempts that study healthcare in federated learning settings. For example, Silva et al. present a federated learning framework by investigating brain structural relationships across diseases and clinical cohorts without sharing individual information \cite{silva2019federated}. Kim et al. convert massive electronic health records into meaningful phenotypes for data analysis by iteratively transferring secure summarized information of hospitals to a central server in federated settings \cite{kim2017federated}. Liu et al. develop a two-stage federated natural language processing method that enables utilization of clinical notes from different hospitals or clinics without moving the data and demonstrate its superior performance \cite{liu2019two}. Besides, healthcare giants around the world --- including the American College of Radiology (ACR), MGH and BWH Center for Clinical Data Science, and UCLA Health --- are piloting FL in healthcare by bringing AI with privacy to hospitals \cite{healthcareIn}. For example, ACR uses FL to allow its medical imaging members to securely build, share and adapt ML models. Partners HealthCare, an integrated health system founded by Brigham and Women's Hospital and Massachusetts General Hospital, has also announced a new initiative using the NVIDIA Clara FL framework that will combine technical and clinical expertise with privacy preservation to deliver more robust AI algorithms for healthcare from hospital to home \cite{MGB}. Nevertheless, these studies of FL for healthcare applications pose new challenges from both statistical and communication perspectives and lacks personalization.

Specifically, in this paper we propose FedHome, a novel framework that aims to achieve accurate in-home health monitoring without compromising user data privacy by leveraging the merits of federated learning and edge computing. As shown in Fig. \ref{framework}, FedHome adopts a cloud-edge architecture for distributed data processing at users' homes, achieves collaborative model training through federated learning and performs personalized model application with user's local data. We elaborate these key elements in detail as follows.

\textbf{Cloud-Edge Architecture.} As elaborated above, purely cloud-based health monitoring would risk at the privacy leakage since it requires users to upload their sensitive health related monitoring data. On the other hand, in some emergency situations (e.g., fall detection for elderly people), the cloud-based approach could fail to respond in real time due to the significant network latency or interruption over the Internet. To address these issues, we promote a synergistic cloud-edge architecture to bring necessary on-demand edge computing power in the proximity of IoT devices. Therefore, each user can choose to process its local data on device, or offload its intensive model training tasks to the trustworthy edge nodes (e.g., edge gateway at home) for fast training. By leveraging federated learning, all users can merge their local models into a global model in the cloud for knowledge sharing without exposing their sensitive data. The trained model deployed locally at the edge can conduct inference much faster than in the cloud, and hence is useful for providing real-time healthcare services. Note that the cloud server here can be provided by some healthcare service company or can be a public server deployed collectively by the entire community of the home users.

Furthermore, as family members in the same home trust each other, they can share data and train a family-shared model at the trustworthy edge nodes, thus family members with insufficient private data can develop accurate local models by reaping the benefits from the data of their mutual-trusting family members.

\textbf{Federated Learning.} The federated learning (FL) procedure of FedHome mainly consists of the following three stages. Indicated by recent work that local models trained from different initial conditions would perform poorly in federated learning \cite{mcmahan2017communication}, we initialize the cloud model and then send the initial state to all clients (homes participated in FL) so that client models can be trained from the same random initialization. After receiving the global cloud model, each client model at the edge then performs local computation based on the global model and its local data, and sends the updated local model parameters to the cloud. Finally, the cloud server aggregates model updates submitted by participant clients and averages these updates into its global model. The process repeats until it converges after many rounds of iterations. Note that all these steps do no share any user data or information but the model parameters.

\textbf{Personalization.} The trained cloud model is based on generic datasets from dispersed homes, which may not well capture the specific characteristics of an individual target user. To perform personalized in-home health monitoring, each user can train a personalized model by integrating the trained global model and her personal health data. Nevertheless, as user's personalized data is usually insufficient and high-skewed, there may be a significant distribution difference between user's local data and the data population of all the participating clients, leading to large model weight divergence between user's personalized model and the global model \cite{zhao2018federated}. Thus, directly retraining the cloud model based on user's local data may result in an even worse model (e.g., degraded accuracy, overfitting with a few data samples) \cite{lin2020real}. To tackle this problem, the model trained in the cloud and the edges should be carefully designed. Specifically, we devise a generative convolutional autoencoder (GCAE) network to deal with the imbalanced class distribution of user's health monitoring data. Besides, GCAE is suitable to be trained under federated learning as it is lightweighted and can reduce the communication cost between the cloud and the edges. The details of GCAE are elaborated in Section \ref{GCAE}.

\section{Learning algorithm for FedHome}
\label{SectionLearning}
In this section, we describe the learning algorithm for FedHome, which consists of personalized federated learning and generative convolutional autoencoder. The key notations used in this paper are summarized in Table \ref{notations}.

\begin{table}[!t]
	\vspace{5pt}
	\caption{List of key notations. }
	\newcommand{\tabincell}[2]{\begin{tabular}{@{}#1@{}}#2\end{tabular}}
	\label{notations}       
	\centering
	\begin{tabular}{|c|c|}
		\hline
		Symbol &  Description\\
		\hline $N$ & Total number of clients in a healthcare system\\
		\hline $K$ & \tabincell{c}{Number of activated clients participating in a communication \\ round} \\
		\hline $\mathcal{D}_{k}$ & Available dataset on client $k$\\
		\hline $D_{k}$ & The set of indexes of  dataset on client $k$\\
		\hline $n$ & \tabincell{c}{Total number of data samples participating  in a \\ communication round}\\
		\hline $n_{k}$ & Number of data samples available from client $k$ \\
		\hline E &  \tabincell{c}{Number of training passes over its local dataset for a\\ client in a communication round}\\
		\hline B &  Local minibatch size used for client updates\\	
		\hline $\eta$ & learning rate of federated learning\\	
		\hline $\lambda$ & weighting factor for the decoder network\\	
		\hline
	\end{tabular}
	\vspace{-10pt}
\end{table}

\begin{figure*}[!t]
	\centering
	\includegraphics[width=0.95\linewidth]{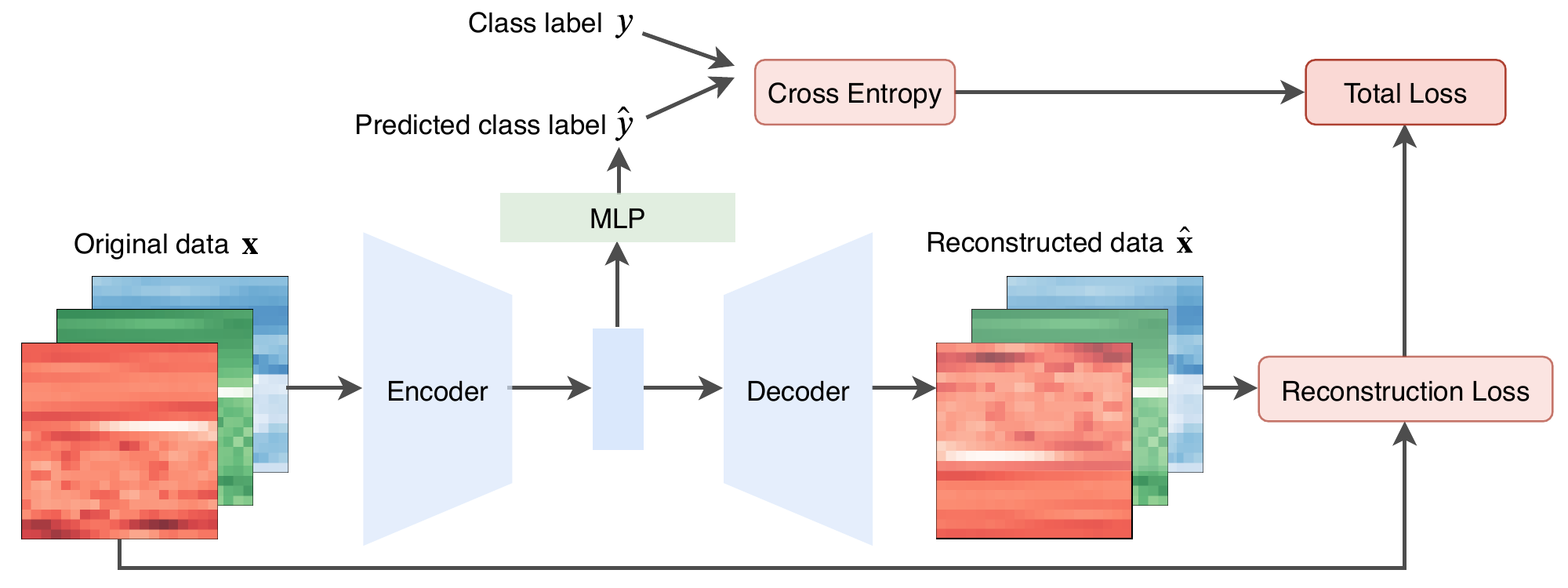}
	\caption{Architecture of the generative convolutional autoencoder.}
	\label{AE_CNN_MLP}
\end{figure*}

\subsection{Personalized Federated Learning}
Federated learning, which enables many participating clients to train a shared global model without sharing their private data, is the core building block for the FedHome framework. In FedHome, both the cloud model and the client models at the edges are learned by deep neural networks (DNNs) as DNNs have been identified as a universal function approximator with excellent generalization and approximation capabilities \cite{hornik1989multilayer}. Let $F_{k}$ and $D_{k}$ denote the local empirical objectives and the set of indexes of local data on client $k$ (edge server $k$), respectively.  $n_{k}$ is the number of samples available from client $k$ and $n = \sum_{k=1}^{K}n_{k}$ is the total sample size in a communication round, where $K$ is the number of active clients participating in federated learning. Thus, federated learning problem boils down to solving an empirical risk minimization problem of the form \cite{mcmahan2017communication}:
\begin{equation}
	\label{objective}
	\min_{\omega \in \mathbb{R}_{d}} f(\omega) = \sum_{k=1}^{K}\frac{n_{k}}{n}F_{k}(\omega) \quad \text{where} \quad F_{k}(\omega) = \frac{1}{n_{k}}\sum_{i\in D_{k}}f_{i}(\omega).
\end{equation}
The objective $f(\omega)$ in federated learning can be rephrased as a linear combination of the local empirical objectives $F_{k}(\omega)$. The learning objective is task-specified, for instance, the objective can be cross-entropy loss for classification tasks. $\omega$ denotes the parameters to be learned, i.e., the weights and bias in our deep neural networks.

\texttt{FederatedAverage} (FedAvg) algorithm, which combines local stochastic gradient descent (SGD) on each client with a server that performs iterative model averaging, serves as the fundamental framework in federated learning settings \cite{mcmahan2017communication}. In each communication round $t$, each edge client $k$ computes $g_{k} = \nabla F_{k}(\omega_{t}) $, the average gradient on its local data at the current model parameters $\omega_{t}$, and then the cloud server aggregates all these gradients submitted from clients and updates the global model by:
\begin{equation}
	\label{update}
	\omega_{t+1} \leftarrow \omega_{t} - \eta \sum_{k=1}^{K} \frac{n_{k}}{n}g_{k},
\end{equation}
since $\sum_{k=1}^{K} \frac{n_{k}}{n}g_{k} = \nabla f(\omega_{t})$. As communication cost is expensive in federated learning, FedAvg proposes to use additional computation in order to decrease the number of communication rounds. Two primary ways to achieve this are outlined as follows: 1) increasing parallel clients participating in each communication round; and 2) increasing computation on each client by iterating the local update $\omega^{k} \leftarrow \omega^{k} - \eta \nabla F_{k}(\omega^{k})$. In each communication round, the amount of computation for each client is controlled by $E$ and $B$, which represent for the number of training passes over its local dataset and the local minibatch size, respectively.

FedAvg has been demonstrated to be accurate and robust in image classification tasks and language modeling tasks \cite{mcmahan2017communication}. In this paper, we adopt FedAvg algorithm for cloud-edge collaborative training in FedHome. Note that, to further enhance the data privacy for the in-home health monitoring scenario, the above parameter exchange and gradient aggregation steps are done with homomorphic encryption \cite{rivest1978data, phong2018privacy}, which is a powerful tool to secure the learning process by computing on encrypted data and has been studied and adopted in FL settings. For example, Federated AI Technology Enabler (FATE) initiated by WeBank's AI division is an open-source technical framework to support federated AI ecosystem that enables secure computing protocols based on homomorphic encryption (HE) and multi-party computation \cite{FATE}. The learning process of FedHealth framework can leverage homomorphic encryption to avoid information leakage of model parameters during the learning process \cite{chen2020fedhealth}. To reduce the computational cost, more efficient variants of HE, such as additive HE, can be adopted in practical FL systems rather than fully homomorphic encryption which is computationally expensive \cite{kairouz2019advances}.

After the global cloud model is learned, it can be directly applied to the homes at the edges for in-home health monitoring. However, it is obvious that the distribution of training samples for the cloud model is highly different from that generated by a single user, thus the cloud model fails in personalization. To make the model more tailored to a specific user, each edge client at home can train a personalized model by integrating the cloud model and the health monitoring data generated by its own users. Nevertheless, as user's personalized data is usually insufficient and high-skewed, there may be a significant distribution difference between user's local data and the data population of all the participating clients, leading to large model weight divergence between user's personalized model and the global model \cite{zhao2018federated}. Thus, directly retraining the cloud model with user's local data may result in an even worse model (e.g., slow convergence rate, degraded accuracy and overfitting with a few data samples) \cite{lin2020real}. To cope with this issue, we devise a generative model called GCAE under federated learning paradigm for better personalization, which is elaborated in the following.

\subsection{Generative Convolutional Autoencoder}
\label{GCAE}

The health monitoring dataset stored in the edge server of home $k$ can be defined as $\mathcal{D}_{k} = \{(\mathbf{x}_{i}, y_{i})\}_{i=1}^{n_{k}}$. The feature $\mathbf{x}_{i}$ contains the health information and is always high-dimensional due to a multitude of monitoring signals for healthcare, and the distribution of class label $y_{i}$ is usually skewed and imbalanced since abnormal states or events such as elderly falling-down are usually sparse during the entire monitoring process. To deal with the imbalanced data distribution issue, we propose to learn a generative prediction model in both the cloud and the edges. More specifically, we devise a generative convolutional autoencoder (GCAE) that utilizes autoencoder (AE) as the backbone. AE aims to learn compact and effective feature representations of input values via an unsupervised learning schema by mapping the input to a latent space through encoder network and then mapping back to the original space using decoder network  \cite{vincent2010stacked}. The learned feature representations are usually low-dimensional and contain all the information needed to recover the original inputs, and thus can be used as feature vector inputs to a supervised learning model (e.g., Multi-Layer Perceptron--MLP) for user health monitoring. It's worthnoting that the feature representations, which are learned by compressing useful information of the local input data into low dimension by the encoder network of GCAE, are always learned and maintained by the local user without submitting to the cloud. And the cloud server only aggregates the model parameter updates obtained by performing local computations based on the global model and the local data of home users, without access to user's sensitive data (e.g., data samples, low-dimensional feature representations), hence user's privacy can be well protected in our FedHome framework.  Besides, due to the fact that health monitoring data is mostly medical images and sensor records, we use convolutional neural network (CNN) as the main architecture for both encoder and the decoder. CNN is competent to extract features from signals and it has achieved promising results in image classification and speech recognition \cite{krizhevsky2017imagenet}.

\begin{figure*}[!t]
	\centering
	\includegraphics[width=0.95\linewidth]{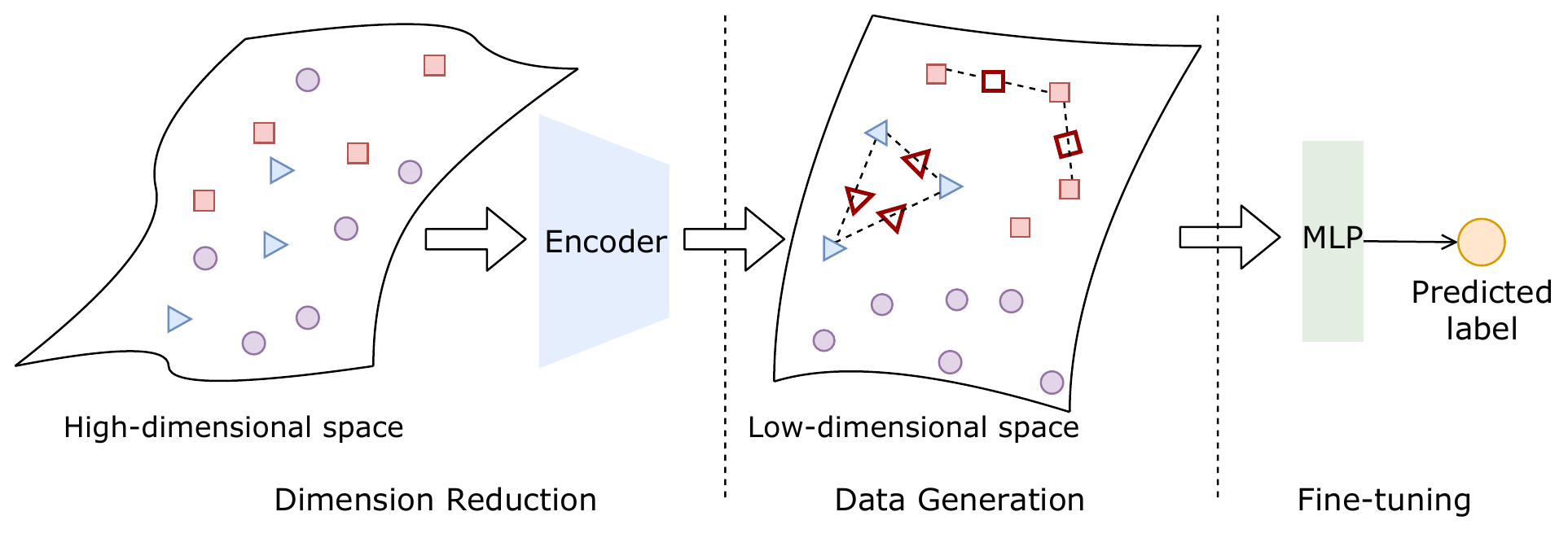}
	\caption{Before inference or classification, the health data is mapped to a low-dimensional space to generate samples in minority classes with SMOTE. With the synthetic class-balanced dataset, we can fine-tune the prediction network for more accurate health monitoring. }
	\label{SMOTE}
\end{figure*}

Fig. \ref{AE_CNN_MLP} depicts the details of GCAE, which consists of a convolutional pyramid encoder $e_{\phi}$ that focuses on learning low-dimensional, common and representative features of high-dimensional inputs $\mathbf{x}$, a convolutional decoder $d_{\theta}$ that calculates the reconstruction loss between $\mathbf{x}_{i}$ and $d_{\theta}(e_{\phi}(\mathbf{x}_{i}))$:
\begin{equation}
	\label{decoder}
	\mathcal{L}_{d}((\phi,\theta),\mathcal{D}_{k}) =  \sum_{i=1}^{n_{k}}||\mathbf{x}_{i}-d_{\theta}(e_{\phi}(\mathbf{x}_{i}))||_{2}^{2},
\end{equation}
and a Multi-Layer Perceptron (MLP) $p_{\psi}$ that predicts the likelihood of the class label $y_{i}$. The objective of MLP is to minimize the difference score between the actual label and predicted probability distributions for all classes:
\begin{equation}
	\label{mlp}
	\mathcal{L}_{p}((\phi,\psi),\mathcal{D}_{k}) = \sum_{i=1}^{n_{k}}loss(y_{i},p_{\psi}(e_{\phi}(\mathbf{x}_{i}))),
\end{equation}
where the loss function $loss(\cdot,\cdot)$ denotes the cross-entropy loss for specific prediction tasks, e.g., human activity recognition task.
As we wish our health monitoring model is equipped with the generation ability, we propose to learn the representative low-dimensional representations and conduct prediction jointly.  Thus, we train an end-to-end model by minimizing:
\begin{equation}
	\label{CAE_target}
	\mathcal{L}(\omega,\mathcal{D}_{k}) = \mathcal{L}_{p} + \lambda \mathcal{L}_{d} ,
\end{equation}
where $\omega = (\phi, \theta, \psi)$ stands for the parameters of the three models and $\lambda \in \mathbb{R}^{+}$ is the weighting factor that controls the impact of the decoder network on the whole model.

In the training process of FedHome, the cloud server managed by smart healthcare service provider first initializes the GCAE model and then sends the initialized model to all the participating clients (e.g., smartphones). Then, each IoT device user can offload its GCAE training task to the trustworthy edge (e.g., smart gateway or edge server) at home for fast computation. Finally, a global GCAE model can be well trained by leveraging model updates from multiple home edges under the coordination of the central cloud server in a cloud-edge architecture.

\textbf{Data Generation.}
When the learned GCAE model is deployed to the edge, we first conduct personalization to make the model more tailored to the edge users. However, the dataset generated by a single user may suffer from data imbalanced problem, which may heavily impact the model performance as a well-trained deep neural network usually relies on a class-balanced training dataset. One common way to tackle the issue of imbalanced data is over-sampling which aims to increase the number of instances from the underrepresented classes in the dataset.

Due to the high dimensional and complex characteristics of health monitoring data, it is hard or even impossible to conduct over-sampling. As encoder network $e_{\phi}$ of GCAE is able to capture representational, common and low-dimensional features of original input $\mathbf{x}_{i}$, we propose to conduct over-sampling in the set of $e_{\phi}(\mathbf{x}_{i})$ instead of $\mathbf{x}_{i}$. In our design, the encoder network of GCAE encodes the high-dimensional input data $\mathbf{x}_{i}$ with dimension of $200 \times (3 + 3) = 1200$ to low-dimension (e.g., 200 in our experiments). Thus, the dimension reduction ratio is 6 in our studying case.  It is flexible to determine the dimension reduction ratio according to the specific healthcare applications. Then we adopt synthetic minority over-sampling technique (SMOTE), one of the most commonly used oversampling methods, to generate a class-balanced dataset \cite{chawla2002smote}. As depicted in Fig. \ref{SMOTE}, SMOTE first randomly selects one or more of the k-nearest neighbors for each instance in the minority classes and then adopts linear interpolation to generate new instances. A salient advantage of over-sampling in low dimension space is that the accuracy in mimicking the distribution of data samples can significantly improve when the dimension reduces \cite{elreedy2019comprehensive}. Thus, a class-balanced dataset in low-dimensional space can be obtained based on a given user's original dataset.

\textbf{Personalization}. As the cloud model only learns the coarse features from all users, we focus on learning the fine-grained information of a particular user based on the reconstructed class-balanced dataset for personalization. Specifically, usually the features in deep networks are highly transferable in lower layers of the network since they focus on learning the common and low-level features, and hence encoder network can be directly reused in the personalized model \cite{yosinski2014transferable}. While the higher layers learn more specific features to the task and the user, which means that we should refine the prediction model $p_{\psi}$ (i.e., Multi-Layer Perceptron--MLP) since it is at the higher layers of GCAE. We fine-tune the model parameters of $p_{\psi}$ by retraining with the reconstructed class-balanced dataset based on user's personal data. In this way, each user can obtain a more accurate personalized model for in-home health monitoring.

\textbf{Communication Overhead Reduction.} GCAE model not only has the ability to generate new data, but also reduces communication overhead to some extent. To well capture the health information and insights behind the data, it is often necessary to train a large neural network which has millions of parameters. However, in federated setting, a lightweight model is highly desirable in order to reduce the communication overhead during model parameter updates among the cloud and the edges.

Parameter sharing scheme is widely-used to reduce the number of training parameters. In our design, both the encoder and decoder networks of GCAE are convolutional neural networks (CNN) in which parameter sharing is used. More concretely, convolutional layer, the core building block of CNN, shares weights by all neurons in a particular feature map (the output received on convolving the image with a particular filter). Max-pooling layer, followed convolutional layer, is to progressively reduce the spatial size of the representation. However, purely adopting CNN will cause a lot of information loss and further lead to inaccurate health monitoring. With the autoencoder mechanism, the encoder network of GCAE is able to learn compact and representative feature vector of user's health monitoring data via unsupervised scheme, ensuring that the low-dimensional input feature vector $e_{\phi}(\mathbf{x}_{i})$ fed in prediction network is nearly lossless. Therefore, GCAE is able to reduce communication overhead without compromising performance, which is of great significance in federated settings.

\begin{algorithm}[htb]
	\caption{The learning procedure of FedHome}
	\label{fedshare}
	\begin{algorithmic}[1]
		\Require Dataset from $N$ distributed clients $\{\mathcal{D}_{1}, \mathcal{D}_{2},\cdots, \mathcal{D}_{N}\}$, participating client number $K$ in each communication round, local minibatch size $B$, number of local epochs $E$, and learning rate $\eta$.
		\State /*federated learning process*/
		\State \textbf{Cloud server executes}:
		\State \quad Construct a GCAE model $\mathcal{M}_{S}(\omega) = \{e(\phi), d(\theta), p(\psi)\}$ as the cloud model and initialize its model parameters $\omega_{0} = \{\phi_{0}, \theta_{0}, \psi_{0}\}$
		\State \quad \textbf{for} each round $t = 0, 1, 2, \cdots$ \textbf{do}
		\State \quad \quad randomly select active client set $U_{t}$ with size $K$ from $N$ clients
		\State \quad \quad distribute $\mathcal{M}_{S}(\omega_{t})$ to all clients in $U_{t}$ via homomorphic encryption
		\State \quad \quad \textbf{for} each client $k \in U_{t}$ \textbf{in parallel do}
		\State \quad \quad \quad $\omega_{t+1}^{k} \leftarrow$ \textbf{ClientUpdate}($k,\omega_{t}$)
		\State \quad \quad \textbf{end for}
		\State \quad \quad upload all client models to the server using homomorphic encryption
		\State \quad \quad /*update cloud model by averaging client models*/
		\State \quad \quad $\omega_{t+1} \leftarrow \sum_{k \in U_{t}}\frac{n_{k}}{n}\omega_{t+1}^{k}$
		\State \quad \textbf{end for}
		\State \quad get the learned global model parameter $\omega_{S}$
		\State \quad distributed the learned model $\mathcal{M}_{S}(\omega_{S})$ to all clients via homomorphic encryption
		\State \quad \textbf{for} each client $k = 1,2, \cdots, N$ \textbf{do}
		\State \quad \quad $\mathcal{M}_{k}(\omega^{k}) \leftarrow$ \textbf{ClientPersonalization}($\mathcal{D}_{k},\omega_{S}$)
		\State \quad \textbf{end for}	
		\Ensure cloud model $\mathcal{M}_{S}(\omega_{S})$ and personalized user model $\mathcal{M}_{k}(\omega^{k})$, $k \in \{1,2, \cdots,N\}$.
	\end{algorithmic}
\end{algorithm}

\begin{algorithm}[htb]
	\caption{Model update on each edge server}
	\label{ClientUpdate}
	\begin{algorithmic}[1]
		\State \textbf{ClientUpdate}($k,\omega$): // Run on client $k$
		\State \quad $\mathcal{B} \leftarrow$ (split $\mathcal{D}_{k}$ into batches of size $B$)
		\State \quad \textbf{for} each local epoch $e$ from $1$ to $E$ \textbf{do}
		\State \quad \quad \textbf{for} batch $b \in \mathcal{B}$ \textbf{do}
		\State \quad \quad \quad $\omega \leftarrow \omega - \eta \nabla \mathcal{L}(w,b)$
		\State \quad \quad \textbf{end for}
		\State \quad \textbf{end for}
		\State \quad return the trained model parameter $\omega$	
	\end{algorithmic}
\end{algorithm}

\begin{algorithm}[htb]
	\caption{Personalization on each edge server}
	\label{ClientPersonalize}
	\begin{algorithmic}[1]
		\State \textbf{ClientPersonalization}($\mathcal{D},\omega$): //Personalization process
		\State \quad calculate low-dimensional feature representation $e_{\phi}(\mathbf{x}_{i})$ for $\mathbf{x}_{i}$ using encoder network of GCAE for all samples in $\mathcal{D}$ and get the low-dimensional dataset $\mathcal{D}^{\prime}$
		\State \quad construct a class-balanced dataset $\mathcal{D}^{\prime}_{b}$ by adopting SMOTE algorithm on $\mathcal{D}^{\prime}$
		\State \quad refine model parameters of prediction network $p(\psi)$ with dataset $\mathcal{D}^{\prime}_{b}$ and get a new prediction model $p(\psi^{\prime})$
		\State \quad return model $\mathcal{M}(\omega^{\prime}) = f(\phi,\theta,\psi^{\prime})$
	\end{algorithmic}
\end{algorithm}

\begin{table*}[!t]
	\normalsize
	\caption{Human activity in MobiAct dataset.}
	\label{ADLsFalls}       
	\newcommand{\tabincell}[2]{\begin{tabular}{@{}#1@{}}#2\end{tabular}}
	\centering
	\begin{tabular}{c|c|c|c}
		\hline Category & Code & Activity & Description \\ \hline
		\multirow{6}{*}{ADLs} & STD & Standing & Standing with subtle movements \\
		&  WAL & Walking & Normal Walking \\
		&  STU & Stairs up & Stairs up (10 stairs) \\
		& STN & Stairs down & Stairs down (10 stairs) \\
		& JUM & Jumping & Continuous jumping \\
		& JOG & Jogging & Jogging \\ \hline
		\multirow{3}{*}{Fall-like Activities} &  CSI & Car step in & Step in a car \\
		& CSO & Car step out & Step out of a car \\
		&SCH & Sit chair & Sitting on a chair \\\hline
		\multirow{4}{*}{Falls} &\multirow{4}{*}{Fall} & Forward-lying & Fall forward from standing, use of hands to dampen fall\\
		& & Front-knees-lying & Fall forward from standing, first impact on knees \\
		& & Sideward-lying & Fall sideward from standing, bending legs \\
		& & Back-sitting-chair & Fall backward while trying to sit on a chair \\ \hline		
	\end{tabular}
\end{table*}
\subsection{The Holistic Algorithm}
To sum up, the holistic mechanism of FedHome is presented in Algorithm \ref{fedshare}, in which the edges train a globally shared GCAE model for in-home health monitoring under the coordination of a cloud server by leveraging federated learning. In each communication round,  each edge uses the current cloud model as the initial model parameters in order to train its local model using its local user data, and then sends its local model parameters to the cloud (see Algorithm \ref{ClientUpdate}). This process repeats until the GCAE model on the cloud converges. Then, the trained GCAE model can be deployed in the edge server of each home. Each home user can further learn a personalized model by synthesizing a class-balanced dataset with her personal data and then refining model parameters with the generated dataset (see Algorithm \ref{ClientPersonalize}). Note that this framework works continuously with the new emerging user data. In other words, FedHome is able to perform incremental learning \cite{rebuffi2017icarl}. That is, when facing new user data, both cloud model and edge client models can be updated continuously. Moreover, the longer the user uses the healthcare application, the more personalized the client model can be. Besides, the learned cloud model, which captures the generic information of healthcare application users, can be easily deployed as a prior model to a new joining home that is new for the healthcare application and has almost no usage records.

The FedHome framework can also adopt other deep neural networks tailored to specific task, for example, recurrent neural networks for modeling Alzheimer's Disease (AD) progression \cite{jung2019unified}, attention network for dementia status prediction from brain magnetic resonance imaging \cite{lian2019end}. This makes FedHome as a general framework flexible for supporting many privacy-preserving healthcare applications.

\begin{figure*}[!t]
	\centering
	\includegraphics[width=0.99\linewidth]{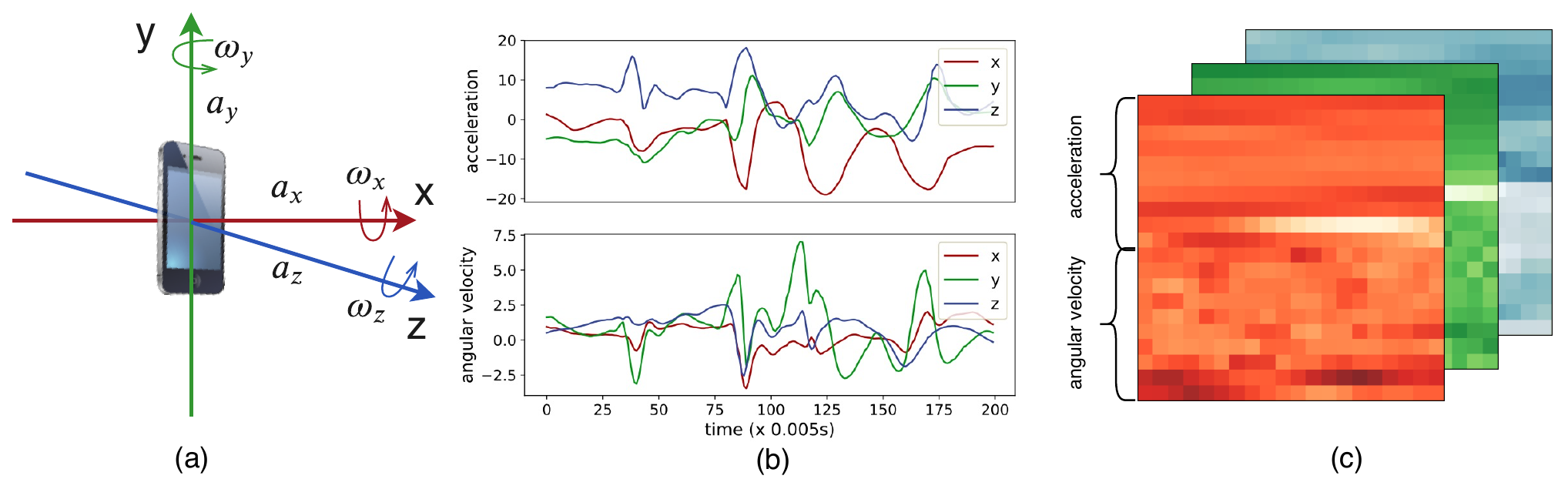}
	\caption{Data preprocessing procedure. (a) Human activity model. (b) Tri-axial acceleration and angular velocity data segmented for one activity. (c) Mapping the tri-axial sensor data into an RGB image.}
	\label{dataPreprocessing}
	\vspace{-10pt}
\end{figure*}

\section{Experiments}
\label{SectionExper}
In this section, we first briefly describe the used dataset and our implementation details for human activity recognition, a specific task highly relevant to in-home health monitoring. Then we evaluate the performance of FedHome framework by comparing with both traditional widely-adopted approaches and federated learning based methods.
\subsection{Dataset Description and Preprocessing}
Our study in this paper is based on a publicly accessible human activity recognition dataset called MobiAct \cite{vavoulas2016mobiact}. There are 57 volunteers (42 men and 15 women) within an age bracket of 20-47 years participating in the generation of MobiAct dataset. Each volunteer wears a Samsung Galaxy S3 smartphone with the accelerometer and gyroscope sensors. The tri-axial linear accelerometer and angular velocity signals are recorded at a constant sampling frequency of 200Hz by the embedded sensors while volunteers perform predefined activities. The recorded activities can be divided into three types: (1) activities of daily living (ADL): the most common everyday activities like walking, standing, stairs up and stairs down; (2) fall-like activities which are sudden or rapid and are similar to falls, such as sitting on a chair or stepping in and out of a car; (3) falls, for example, fall forward from standing, fall backward while trying to sit on a chair, etc. Typically, there are four different types of falls as described in Table \ref{ADLsFalls}. Note that this dataset can provide the relevant application scenario to mimic in-home health monitoring through human activity recognition (e.g., fall detection for elderly people).

Before data preprocessing, we first introduce human activity model which is established based on the Cartesian coordinate system \cite{grood1983a} in accordance with the direction of the sensors in a smartphone as shown in Fig. \ref{dataPreprocessing} (a). In the 3-dimensional coordinate system, $a_{x}$, $a_{y}$, $a_{z}$ denote the acceleration along the $x$, $y$ and $z$ axis, respectively. Correspondingly, $\omega_{x}$, $\omega_{y}$, $\omega_{z}$ represent for the angular velocity of the human body around the $x$, $y$ and $z$ axis. Since one second is enough for users to perform an activity ( e.g., from falling down to touching the ground), a fixed 1-second sliding window is used for feature extraction and the overlapping rate between successive windows is $80\%$. Fig. \ref{dataPreprocessing} (b) shows the tri-axial acceleration and angular velocity data segmented for one activity. As the sampling frequency is 200 Hz, each activity contains 200 data points for each sensor along each axis, and totally $200 \times (3 + 3) = 1200$ data points. If we consider the 3-axes of the human activity model as the 3-channels of an RGB image, the values of the tri-axial acceleration and angular velocity data can be mapped into the pixels in an RGB image. As depicted in Fig. \ref{dataPreprocessing} (c), the first 200 data in the RGB image are tri-axial accelerations while the last 200 data are tri-axial angular velocities.

To practically mimic smart healthcare environments, we randomly select 30 volunteers and regard them as users in multiple dispersed homes. Each user can offload their model training task to the trustworthy edge node (e.g., smart gateway or personal computer at home) for fast training or train the model locally on device. Moreover, as the skewed data distribution is an inherent characteristic in federated settings, we employ three approaches of training data partition:
\begin{itemize}
	\item \textbf{Balanced data partition:} There are 30 homes with one isolated user at each home. For each user, we extract 48 samples for each activity. Totally, one user has 480 samples for model training.
	\item \textbf{Imbalanced data partition:} There are 30 homes with one isolated user at each home. For each user, we take a random number of samples for each activity and finally, we ensure that each user has 480 samples for model training.
	\item \textbf{Home data partition:} The dataset for each user is the same with imbalanced data partition case. However, we randomly divide these 30 users into 10 homes with 1 to 5 family members per home.  Users can share their health data with their family members at the edge.
\end{itemize}
The test data for the above three data partition manners is the same. For the selected 30 users, each user has 160 samples under balanced distribution. It is worth noting that the data distribution is always non-IID as different users have different physical characteristics, lifestyles and lifelong medical data, and as a consequence, different data distributions.

\begin{figure*}[!t]
	\centering
	\subfigure[Test Accuracy]{
		\includegraphics[width=0.455\linewidth]{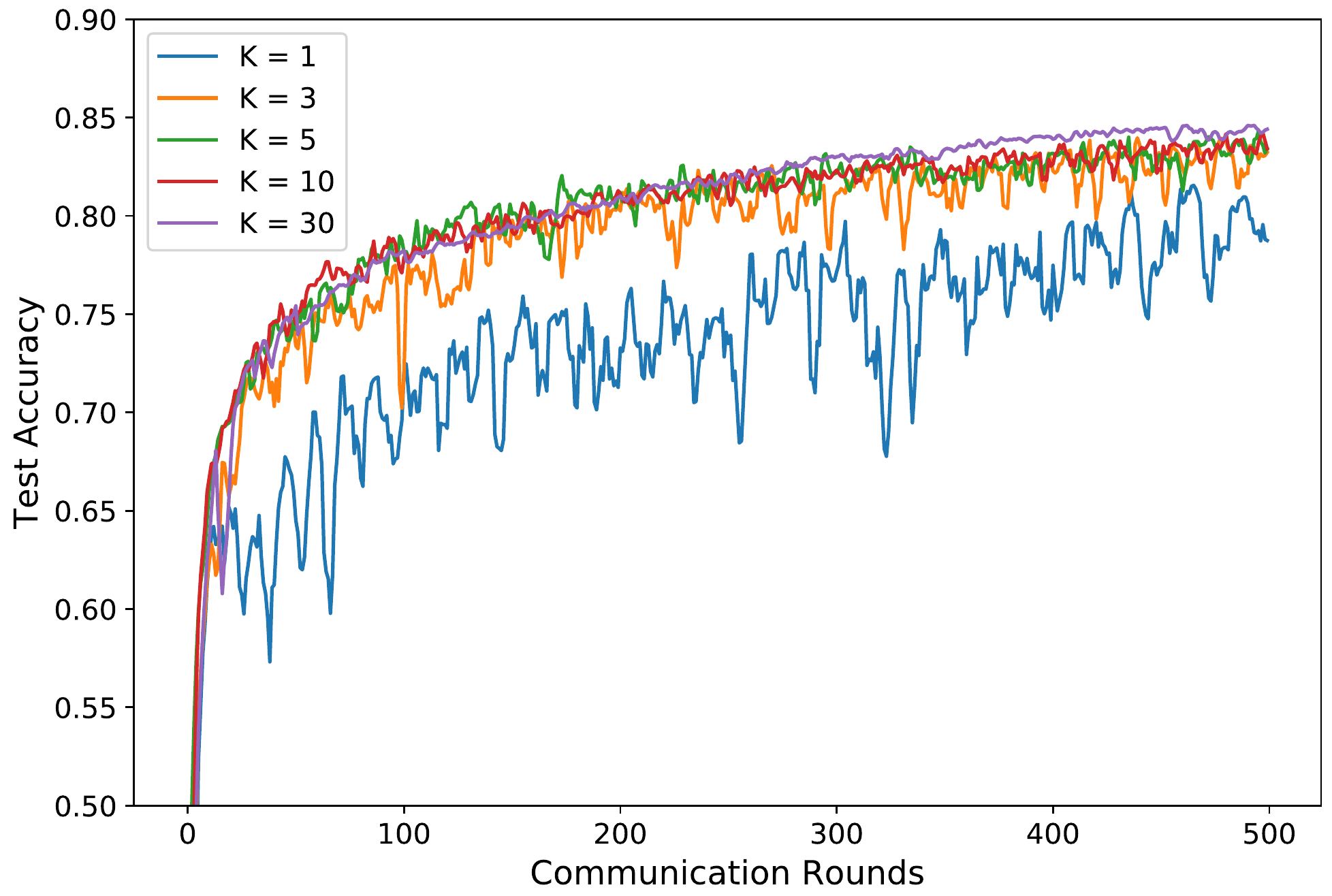}
	}
	\subfigure[Time Cost]{
		\includegraphics[width=0.455\linewidth]{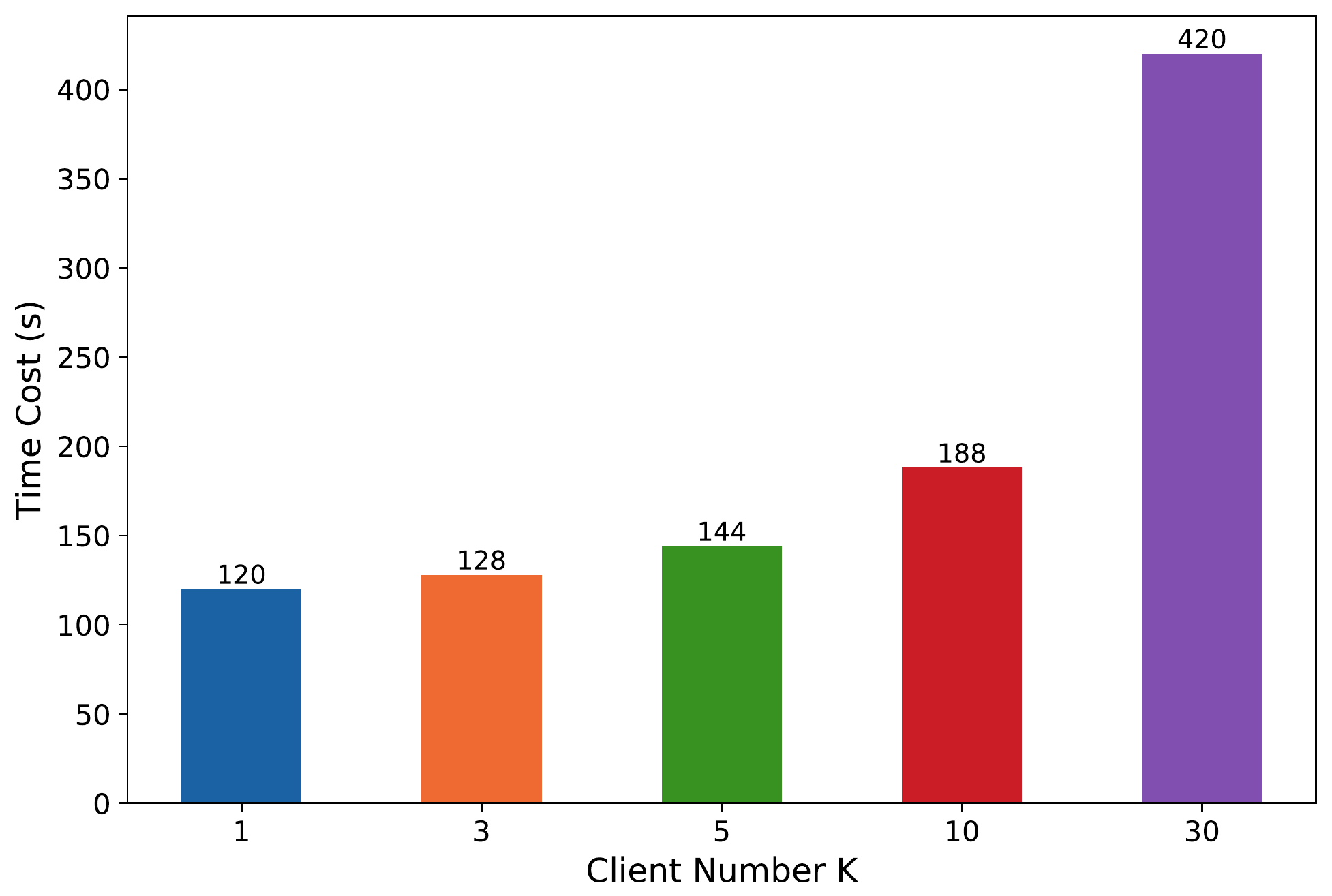}
	} 	
	\caption{The test accuracy and time cost under different number of participating clients in each communication round. We choose K = 5 by making a trade off between the stability and the efficiency of the learning algorithm.}
	\label{KBE}
	\vspace{-7pt}
\end{figure*}

\subsection{Implementation Details}
Generative convolutional autoencoder (GCAE) network is adopted on both the cloud server and the edges at users' homes for model training and prediction. The encoder network of GCAE is composed of 3 convolutional layers with filter size of $3 \times 3$ to encode the original data into a hidden latent representation. The filter numbers of the three convolutional layers are 32, 16 and 8, respectively, and each of the first two convolutional layers is followed by a $2 \times 2$ max-pooling layer. For the decoder network, three $5 \times 5$ deconvolution layers with 16, 32, 3 filters and two $2 \times 2$ up-sampling layers are applied to decode the hidden representation. The activation function for the third convolutional layer is $Sigmoid$ and we use $Relu$ function for other layers in both the encoder and the decoder networks. The output of the encoder is flattened and then fed in an MLP network which is composed of 2 fully-connected layers with 128 and 10 units, respectively. $Softmax$ is adopted in the final fully-connected layer of the MLP network to calculate the probability of the output results. The weighting factor of the decoder network, $\lambda$, is set to be $0.01$. The model is trained by minibatch Stochastic Gradient Descent (SGD) optimizer with a learning rate of 0.01. These hyperparameters are tuned using cross-validation procedure.

\subsection{Experimental Results}
\subsubsection{\textbf{Baselines}} The baselines for human activity recognition can be divided into the following two groups:

\textbf{Traditional centralized algorithms}: traditional algorithms usually collect a large amount of user data in a centralized cloud server to train a satisfactory model. Support Vector Machine (SVM), k-Nearest Neighbor (kNN) and Random Forest (RF) are widely adopted in many healthcare applications \cite{ward2016towards}. With the recent success of deep learning, Multi-Layer Perceptron (MLP) and Convolutional Neural Network (CNN) are also introduced to in-home health monitoring and have achieved great performance improvement \cite{wang2019deep}. We also test the accuracy of Generative Convolutional Autoencoder (GCAE) proposed in this paper in a centralized fashion. We collect all the training data of 30 users to form the training dataset of traditional centralized algorithms and describe these algorithms as below.

\begin{itemize}
	\item \textbf{Support Vector Machine (SVM)} is a popular classifier that is demonstrated to be effective on a wide range of classification problems \cite{shi2009mobile}.
	\item \textbf{k-Nearest Neighbor (kNN)} is a non-parametric method. It is introduced to measure the difference or similarity between instances according to a distance function \cite{he2016smart}.
	\item \textbf{Random Forest (RF)} is an ensemble learning method that builds a set of decision trees with random subsets of attributes and then bags them for classification results \cite{yuan2014power}.
	\item \textbf{Multi-Layer Perceptron (MLP)} is a supervised learning technique that makes final predictions through fully-connected layers. In our design, the MLP network is composed of three fully-connected layers with 1200, 100 and 10 neural units.
	\item \textbf{CNN} has strong representation capability and is widely adopted in healthcare scenarios, such as in human activity recognition \cite{he2019low}. CNN used in this paper is designed as the combination of an encoder and an MLP classifier which has the same architecture with that in our GCAE model.
	\item \textbf{Generative Convolution Autoencoder (GCAE)} takes advantage of the feature extraction ability of CNN, dimension reduction capability of autoencoder and over-sampling strategy of SMOTE \cite{chen2017deep}.
\end{itemize}

\begin{table*}[!t]
	\normalsize
	\caption{Test accuracy of the models in human activity recognition. }
	\label{acc_compare}       
	\newcommand{\tabincell}[2]{\begin{tabular}{@{}#1@{}}#2\end{tabular}}
	\centering
	\begin{tabular}{|c|c|c|c|}
		\hline
		\multirow{2}{*}{Methods} & \multicolumn{2}{c|}{Test Accuracy}  & \multirow{2}{*}{Model Parameters} \\
		\cline{2-3}
		&  Balanced Data & Imbalanced Data & \\	\hline
		SVM & $77.25 \pm 1.77\%$ & $67.88 \pm 0.50\%$ & $-$ \\\hline
		KNN & $80.85 \pm 1.48\%$ & $74.99 \pm 0.67\%$ & $-$ \\\hline
		RF & $84.27 \pm 0.34\%$ & $74.28 \pm 0.29\%$ & $-$ \\\hline
		MLP & $92.31 \pm 1.45\%$ & $87.94 \pm 1.32\%$ & $-$ \\\hline
		CNN & $91.77 \pm 1.43\%$ & $87.92 \pm 0.15\% $ & $-$ \\\hline
		GCAE & $92.02 \pm 0.92 \%$ & $88.10 \pm 0.14\% $  & $-$ \\\hline\hline
		FL-MLP & $89.28 \pm 1.17 \%$ & $85.06 \pm 0.59 \%$ & $1,562,310 \, (29.96 \times)$ \\\hline
		FL-CNN & $85.07 \pm 0.44 \%$ & $82.91 \pm 0.34\% $  & $33,698 \, (0.65 \times)$ \\\hline
		FL-CNN-Large & $87.24 \pm 0.46 \%$ & $84.87 \pm 0.72\% $  & $77,498 \, (1.49 \times)$ \\\hline
		FedHome-p & $89.13 \pm 0.93 \%$  & $84.22 \pm 0.37 \%$ & $52,149$ \\\hline
		FedHome & $\mathbf{95.87 \pm 0.23\%}$ & $\mathbf{95.41 \pm 0.11\%}$ & $\mathbf{52,149}$ \\
		\hline
	\end{tabular}
\end{table*}

\textbf{Federated learning based methods}: in federated settings, each client trains a local model with its personal-generated data, and FedAvg method is applied to obtain a globally shared model by aggregating local models without compromising users' privacy. This process repeats until it converges.
\begin{itemize}
	\item \textbf{FL-MLP:} both the cloud and client models are implemented with MLP.
	\item \textbf{FL-CNN:} CNN model is adopted as both the cloud and client models.
	\item \textbf{FL-CNN-Large:} The difference between FL-CNN-Large and FL-CNN is that the filter numbers of the three convolutional layers in the CNN model of FL-CNN-Large are  64, 32 and 16, respectively.
	\item \textbf{FedHome:} FedHome is the proposed federated learning framework for in-home health monitoring. GCAE is used as both the cloud and the client models. When the training process converges, the client will first balance its data distribution and then conduct model personalization with its own personal data.
	\item \textbf{FedHome-p:} FedHome framework without client personalization is named FedHome-p.
\end{itemize}

\begin{figure}[!t]
	\centering
	\includegraphics[width=0.98\linewidth]{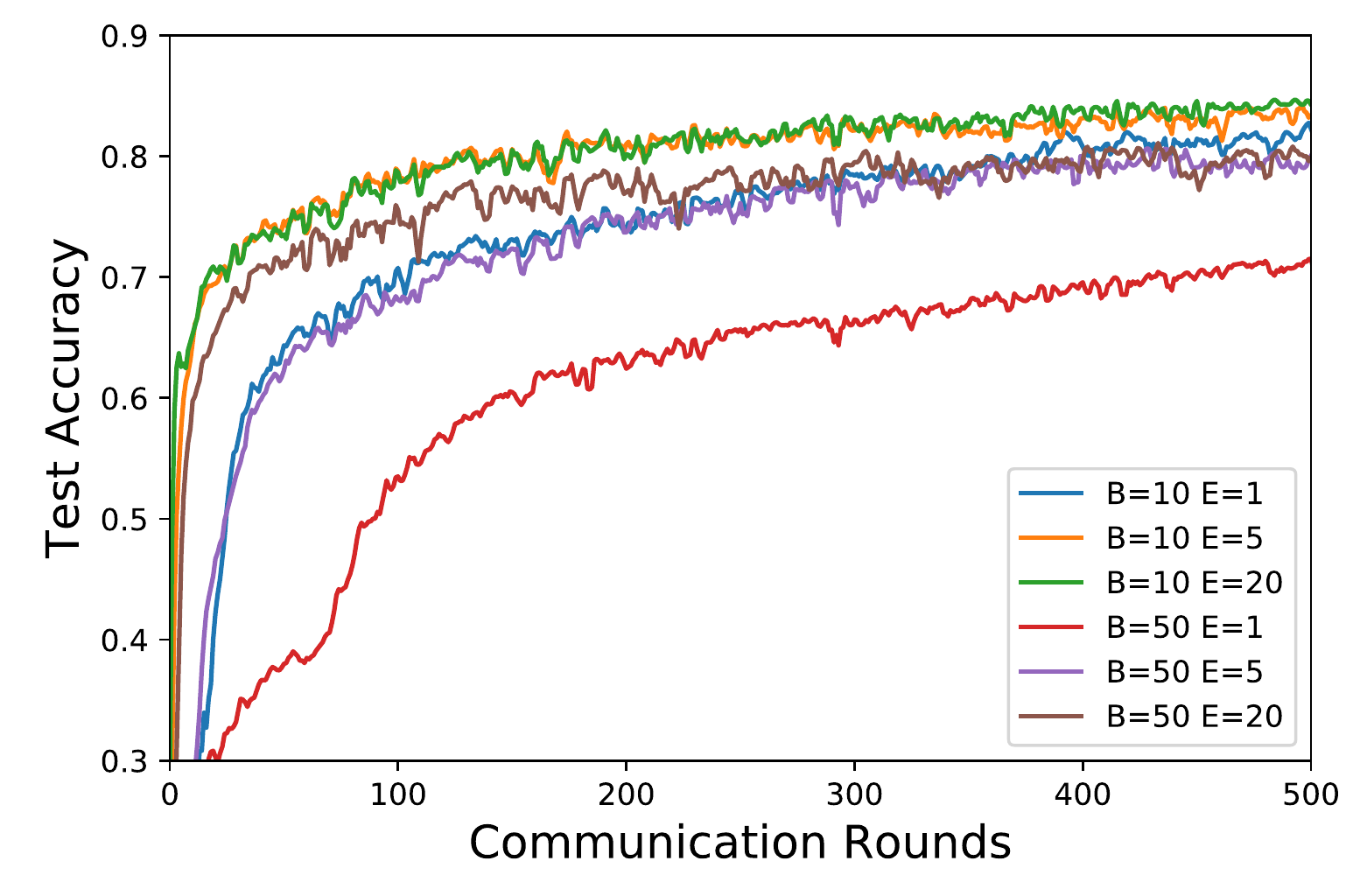}
	\caption{Test accuracy vs. communication rounds under different combinations of $B$ and $E$.}
	\label{BE}
	\vspace{-10pt}
\end{figure}

\subsubsection{\textbf{Performance Evaluation}}
In federated settings, there are three key parameters which may influence the performance of federated learning: (1) $K$, the number of participating clients (homes participating in FL) in each communication round; (2) $E$, training passes over each client's local data and (3) $B$, local minibatch size. We first conduct experiments with the number of clients participating in each communication round. First, we fix $B = 10$ and $E = 5$, and evaluate the test accuracy under different values of $K$. More concretely, we study the effect of $K$ by setting $K$ equal to 1, 3, 5, 10 and 30, which can also be regarded as $\frac{1}{30}$, $\frac{1}{10}$, $\frac{1}{6}$, $\frac{1}{3}$ and $100\%$ of the whole training data. As shown in Fig. \ref{KBE} (a), the learning algorithms with different choices of $K$ can all converge within 500 communication rounds. The test accuracies under different values of $K$ improve with the increase of the communication rounds. However, when $K$ is small (e.g., $K = 1$), the test accuracy shows a significant gap compared with the case with bigger $K$. For example, the test accuracy is $79.15\%$ when $K = 1$ while $84.23\%$ in the case that $K = 30$. Besides, the learning curve exists erratic fluctuation to some extent with a smaller $K$ and becomes smoother as $K$ increases. Although the test accuracies are similar except the case that $K = 1$, the training time for each value of $K$ varies dramatically as demonstrated in Fig. \ref{KBE} (b). For example, the training time in the $K=1$ case is 120s when experimenting on a desktop PC equipped with a quad-core Intel processor at 3.4GHz with 8GB of RAM. The training time for $K = 3$ is $1.07$ times longer compared with the training time when $K = 1$, while the training time for $K = 30$ is $3.28$ times more than that in the $K = 3$ case. We make a trade-off between the stability and the efficiency of the learning algorithm in the training process and fix $K = 5$ for the following experiments.

To unfold the impact of parameter $B$ and $E$ on test accuracy and communication cost, we set six combinations of parameter settings as shown in Fig. \ref{BE}. We can see that the learning curve converges rapidly with larger $E$ and smaller $B$. When $B = 10$ and $E = 20$, the test accuracy of human activity recognition can reach $83.57\%$ within only 300 communication rounds. While for the case that $B = 50$ and $E = 1$, it may require a large number of communication rounds to reach the same test accuracy. In fact, for a client with $n_{k}$ local samples, the number of local updates per round is given by $E\frac{n_{k}}{B}$. Thus, either increasing $E$, decreasing $B$, or both, will result in more computation per client on each round. This observation inspires us that adding more local SGD updates per communication round can produce a dramatic decrease in communication cost, which is essential in federated learning. As the performance of the case  $B = 10$ and $E = 5$ is comparable with the best one, we choose to use $B = 10$ and $E = 5$ under the consideration of computation cost for the following experiments. It's worthnoting that the training process of FedHome can be conducted offline to learn a satisfactory shared model by reaping the benefits of user data from multiple homes without compromising their data privacy. When the globally shared model is deployed on edges or user's IoT devices, FedHome enables fast personalization (retraining with local data of a family or a user) and real-time inference (prediction with the input data by simply conducting matrix multiplications of the local neural network model).

Table \ref{acc_compare} illustrates the performances of different human activity recognition approaches under both balanced and imbalanced data cases. For each method, we compute the average and standard deviation of test accuracy by repeating the training and prediction processes five times. We can see that the test accuracies of all methods in balanced dataset are higher than those in imbalanced dataset. Moreover, the results show that deep learning based methods (MLP, CNN and GCAE) can all achieve a higher accuracy than traditional machine learning methods (SVM, KNN and RF) when the data samples of all clients are aggregated in a centralized location. This is due to the strong representation capability of deep neural networks. Besides, deep learning based methods can be updated and enhanced by incremental learning without retraining. This property is valuable in federated learning where model reuse is important and helpful because new data emerges continuously and the model should be quickly adapted to it. When adopting these deep learning methods in federated settings, the test accuracy decreases to some extent. For example, the test accuracy of FL-MLP under balanced data case is $3.03 \%$ smaller than that of MLP approach, this is mainly because of the non-IID nature of datasets generated by the users at different homes. The imbalanced data samples for each user or home will also deteriorate the test accuracy. Nevertheless, our proposed FedHome approach can mitigate the statistical challenge inherent in federated settings (imbalanced and non-IID nature of dataset from different users) and improve the test accuracy to $95.87\%$ for balanced data case and $95.41\%$ for the imbalanced data case. Specifically, we also examine the performance of FedHome without personalization, which is essentially a federated version of GCAE. The results show that test accuracy will drop by $6.74\%$ and $11.19\%$ in balanced and imbalanced data cases, respectively, when compared with FedHome.

\begin{table}[!t]
	\normalsize
	\caption{The average precision, recall and f1-score for each activity. }
	\label{PRF}       
	\centering
	\begin{tabular}{|c|c|c|c|}
		\hline
		Activity &  Precision & Recall & F1-score \\
		\hline Fall &  96.24\% & 90.2\% & 93.10\% \\
		\hline SCH &  96.20\% & 99.54\% & 97.87\% \\
		\hline CSI &  93.54\% & 90.00\% & 91.30\% \\
		\hline CSO &  93.40\% & 93.70\% & 93.47\% \\
		\hline STU &  95.23\% & 92.63\% & 93.83\% \\
		\hline STN &  93.60\% & 93.97\% & 93.70\% \\
		\hline JUM &  99.01\% & 98.83\% & 98.97\% \\
		\hline JOG &  96.44\% & 98.87\% & 97.63\% \\
		\hline STD & 99.60\% & 99.30\% & 99.50\% \\
		\hline WAL &  95.40\% & 98.20\% & 96.63\% \\ \hline
		
	\end{tabular}
	\vspace{-5pt}
\end{table}

As for communication efficiency which is essential in federated learning, we give the number of model parameters of all the FL based approaches in Table \ref{acc_compare}. Our proposed FedHome framework only has $52,149$ model parameters and will reduce the communication load of model transfer in each round significantly compared with FL-MLP whose model parameter number is nearly $30$ times more than that of FedHome. Although FL-CNN has fewer parameters than FedHome, it yields 10.8-12.5\% accuracy degradation. As scaling up CNN size (e.g., width, depth, etc.) is known to be an effective approach for improving model accuracy \cite{he2020group}, we also compare our FedHome model with FL-CNN-Large, a CNN-based model whose parameter number is 1.49 times larger than that of FedHome. The experimental results show that FedHome can significantly outperform FL-CNN-Large in both balanced and imbalanced data partitions. Moreover, even FedHome-p, a federated version of GCAE without personalization, can achieve comparable performance to FL-CNN-Large, indicating that GCAE can reduce the communication overhead without sacrificing model accuracy. In short, our experiments demonstrate the effectiveness and applicability of our proposed FedHome framework in realistic computation and communication-constrained in-home health monitoring applications.

To give a more detailed performance evaluation, we also show the average precision, recall and f1-score over all 30 users for each activity. As depicted in Table \ref{PRF}, FedHome achieves excellent performance on all the metrics for all activities. For example, the precision scores for JUM and STD activities are higher than $99\%$, which means that FedHome can accurately identify these types of activities. Even for fall and fall-like activities, the precision and recall scores are all over $90\%$, indicating that our proposed FedHome framework is able to well distinguish falls from similar activities. Thus, for some healthcare applications which focus on fall detection, FedHome can eliminate the risk of fall-induced injuries among the elderly by signaling for help when detecting falls.
\begin{figure}[!t]
	\centering
	\includegraphics[width=0.95\linewidth]{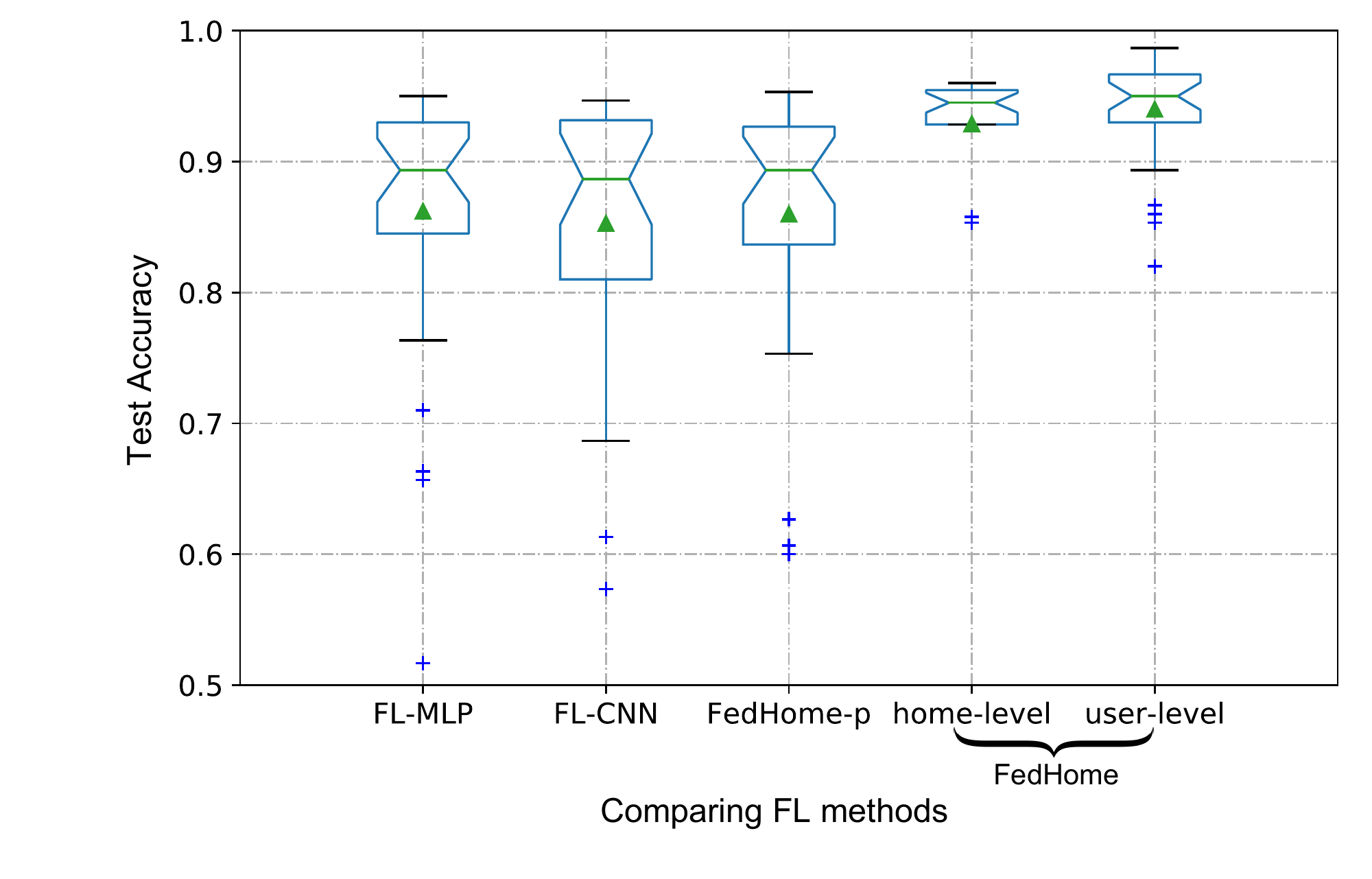}
	\caption{Test accuracy of different FL approaches under home data partition.}
	\label{acc_box}
	\vspace{-10pt}
\end{figure}

For home data partition, since the training data setting is the same with imbalanced data partition when applied to centralized methods, we only conduct experiments on federated learning approaches. As shown in Fig. \ref{acc_box}, all approaches can achieve a high average accuracy of 30 users. However, the test accuracies for these 30 users predicted by FL-CNN vary dramatically. For example, the accuracy of some users may lower than $70\%$ while some users can reach a high accuracy of more than $95\%$. By integrating representation ability of AE which can reduce information loss to some extent with parameter sharing, FedHome-p approach can enhance the prediction performance comparing with FL-CNN. Although FL-MLP can achieve more competitive performance than FedHome-p, it is rarely applicable to realistic applications due to its large training parameter size and high communication cost. As family members in the same home trust and share data with each other, they can choose to perform personalization at either home level or user level. For example, users who have insufficient private data to develop accurate local models can choose home-level personalization in order to benefit from their mutual-trusting family members. While users with abundant data samples can conduct user-level personalization with their personal data. The average test accuracies of FedHome personalized at home level and user level are $92.97\%$ and $94.03\%$, respectively. Moreover, we can see that the test accuracies of 30 users vary in a very small scale when performing home-level personalization. This is largely due to the fact that users with highly-skewed personal data can utilize their family members' health data to enhance their own models. While for some users, personalizing their model using their own data can make the refined model more tailored to them.

\begin{figure}[!t]
	\centering
	\includegraphics[width=0.98\linewidth]{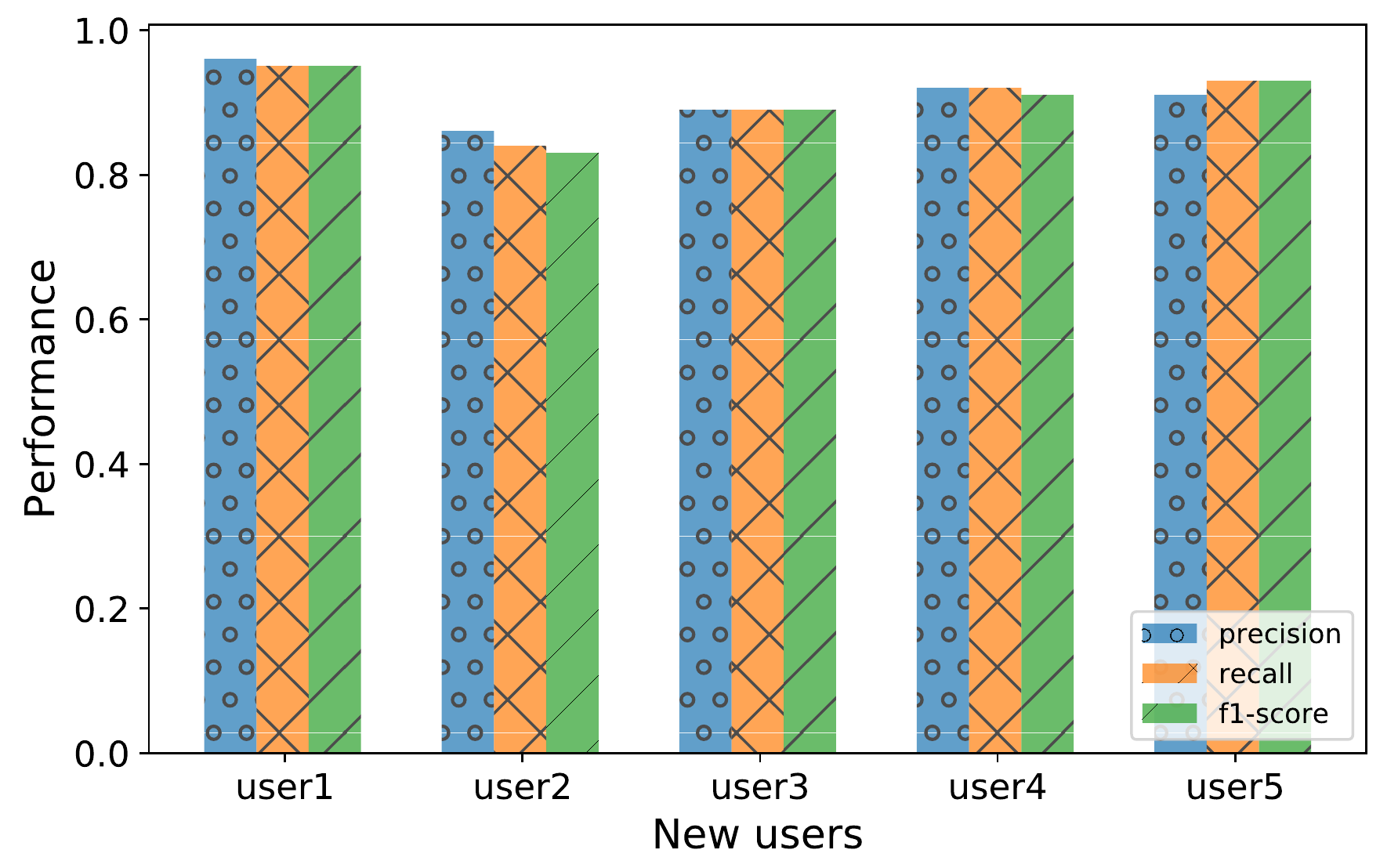}
	\caption{Performance of FedHome framework when facing new users.}
	\label{newUsers}
	\vspace{-10pt}
\end{figure}

We also investigate the generalization ability of FedHome when facing new users. We first randomly select 5 users who have not participated in the cloud model training, and then distribute the learned cloud model to these 5 users. After training with only a few rounds using user's personal data, the local model for each user can achieve a good performance. For instance, as depicted in Fig. \ref{newUsers}, the precision score of user 1 is $96\%$ and the average precision score of all 5 users is $90.8\%$.

\section{Related Work}
\label{RelatedWork}
In this section, we elaborate the recent works in in-home health monitoring and federated learning.
\subsection{In-Home Health Monitoring}
In-home health monitoring improves healthcare effectiveness and lowers healthcare costs by utilizing the mobile and ambient sensors in smart homes, such as smartphones and smartwatches. Due to its importance and urgency in the context of ageing population worldwide, smart healthcare has evoked notable scientific interest \cite{cook2018using}. For example, continuous monitoring and automatic classification of tremor severity in Parkinson's Disease are conducted using a wrist-watch-type wearable device consisting of an accelerometer and a gyroscope \cite{jeon2017automatic}. Physiological signals (e.g., electroencephalography) are used to detect certain diseases, such as seizures \cite{menshawy2015automatic}. Besides, many studies focus on sensor-based human activity recognition and fall detection due to their wide range of application domains including assisted living, sport, human-computer interaction and healthcare \cite{chikhaoui2018cnn, wang2019deep}. In these studies, machine learning approaches, such as support vector machine (SVM), random forest (RF), are widely used for health status monitoring \cite{ward2016towards}. With the recent success of deep learning, many healthcare applications adopt some representative networks including convolutional neural network (CNN), autoencoder (AE), and have achieved satisfactory performances \cite{wang2019deep}. It is noteworthy that these traditional health monitoring applications often build models by aggregating all the user data in a centralized location and lack privacy protection and personalization.

Muhammen et al. propose a personalized ubiquitous cloud and edge-enabled networked healthcare system called UbeHealth \cite{muhammed2018ubehealth}. However, they only emphasize that security policies should be developed and maintained from the device manufacturers, software developing organizations and regulators that check the standard and safety, without protecting user privacy security at the algorithm level \cite{sametinger2015security}. Zhang et al. propose a collaborative cloud-edge computation system  for personalized driving behavior modeling \cite{zhang2019collaborative}. Although this method protects user privacy by keeping user data at the edge devices, it requires a large amount of data that are anonymized and integrated from users who would like to share their data in the cloud, which is impractical in real-world applications.

\subsection{Federated Learning}
Federated learning is first proposed by Google to solve the data island and user privacy problems \cite{konevcny2016federated}. The key idea is to build machine learning models based on datasets distributed across multiple devices while preventing data leakage \cite{yang2019federated}. McMahan et al. introduce the \texttt{FederatedAverage} (FedAvg) algorithm which has been verified to be robust and has been regarded as the fundamental framework in federated learning settings \cite{mcmahan2017communication}. While solving the problems of data island and privacy security, federated learning (FL) poses new challenges, such as imbalanced and highly-skewed data distribution and high communication cost. Thus, recent improvements have been focusing on overcoming these statistical and communication challenges \cite{smith2017federated, konevcny2016federated, lin2017deep}. Zhao et al. propose a data-sharing strategy to improve FedAvg with non-IID by creating a small subset of data which is globally shared among clients \cite{zhao2018federated}. To reduce communication overhead, Jeong et al. propose federated distillation, a distributed knowledge distillation method whose communication payload size depends not on the model size but the output dimension \cite{jeong2018communication}. Prior to operating federated distillation, the authors empower each participating device to locally reproduce the data samples of all devices so as to make the training dataset become IID by using a trained generative adversarial network. Nevertheless, this approach requires the user device to upload a few seed data samples to the server, which may put user privacy as risk. There are also some research efforts targeting at making federated learning more personalizable \cite{chen2018federated}. For example, the proposed FedHealth framework achieves personalized model learning for each organization through knowledge transfer while ignoring the fact that class imbalance problem will bring huge prediction errors \cite{chen2020fedhealth}.

\section{Conclusion}
\label{SectionConclusion}
In this paper, we propose FedHome, a novel cloud-edge federated learning framework for personalized in-home health monitoring, which achieves privacy protection by keeping user data locally. FedHome aggregates the data from multiple homes and trains a global model without compromising user privacy, and then achieves personalized model learning through knowledge transfer. To tackle the statistical and communication challenge inherent in federated learning, the cloud model to be learned is designed as a generative convolutional autoencoder (GCAE), which empowers to synthesize samples of minority classes and form a class-balanced dataset in personalization procedure to address the imbalanced and non-IID data dilemma. Moreover, GCAE consists only a small number of model parameters, and thus can significantly reduce the communication overhead during model transfer. Extensive experiments on human activity recognition have demonstrated the effectiveness of the proposed framework in both evaluation performance and communication efficiency. FedHome can be applied to many healthcare applications without incurring data leakage and can be a powerful approach for in-home health monitoring in the future.


%

\ifCLASSOPTIONcaptionsoff
\newpage
\fi



%
\bibliographystyle{unsrt}
\bibliography{reference}

\begin{thebibliography}{10}

\bibitem{WHO}
World Health Organization.~Towards policy~for health and ageing.
\newblock \url{http:www.who.int/ageing/publications/alc_fs_ageing_policy.pdf}.
\newblock accessed: 2019-12-15.

\bibitem{mano2016exploiting}
Leandro~Y Mano, Bruno~S Fai{\c{c}}al, Luis~HV Nakamura, Pedro~H Gomes,
  Giampaolo~L Libralon, Rodolfo~I Meneguete, P~Rocha Geraldo~Filho, Gabriel~T
  Giancristofaro, Gustavo Pessin, Bhaskar Krishnamachari, et~al.
\newblock Exploiting iot technologies for enhancing health smart homes through
  patient identification and emotion recognition.
\newblock {\em Computer Communications}, 89:178--190, 2016.

\bibitem{verma2018fog}
Prabal Verma and Sandeep~K Sood.
\newblock Fog assisted-iot enabled patient health monitoring in smart homes.
\newblock {\em IEEE Internet of Things Journal}, 5(3):1789--1796, 2018.

\bibitem{hiremath2014wearable}
Shivayogi Hiremath, Geng Yang, and Kunal Mankodiya.
\newblock Wearable internet of things: Concept, architectural components and
  promises for person-centered healthcare.
\newblock In {\em 2014 4th International Conference on Wireless Mobile
  Communication and Healthcare-Transforming Healthcare Through Innovations in
  Mobile and Wireless Technologies (MOBIHEALTH)}, pages 304--307. IEEE, 2014.

\bibitem{voigt2017eu}
Paul Voigt and Axel Von~dem Bussche.
\newblock The eu general data protection regulation (gdpr).
\newblock {\em A Practical Guide, 1st Ed., Cham: Springer International
  Publishing}, 2017.

\bibitem{inkster2018china}
Nigel Inkster.
\newblock {\em China’s Cyber Power}.
\newblock Routledge, 2018.

\bibitem{xu2020federated}
Jie Xu, Benjamin~S Glicksberg, Chang Su, Peter Walker, Jiang Bian, and Fei
  Wang.
\newblock Federated learning for healthcare informatics.
\newblock {\em Journal of Healthcare Informatics Research}, pages 1--19, 2020.

\bibitem{lim2020federated}
Wei Yang~Bryan Lim, Nguyen~Cong Luong, Dinh~Thai Hoang, Yutao Jiao, Ying-Chang
  Liang, Qiang Yang, Dusit Niyato, and Chunyan Miao.
\newblock Federated learning in mobile edge networks: A comprehensive survey.
\newblock {\em IEEE Communications Surveys \& Tutorials}, 2020.

\bibitem{zhao2018federated}
Yue Zhao, Meng Li, Liangzhen Lai, Naveen Suda, Damon Civin, and Vikas Chandra.
\newblock Federated learning with non-iid data.
\newblock {\em arXiv preprint arXiv:1806.00582}, 2018.

\bibitem{jeong2018communication}
Eunjeong Jeong, Seungeun Oh, Hyesung Kim, Jihong Park, Mehdi Bennis, and
  Seong-Lyun Kim.
\newblock Communication-efficient on-device machine learning: Federated
  distillation and augmentation under non-iid private data.
\newblock {\em arXiv preprint arXiv:1811.11479}, 2018.

\bibitem{chen2020fedhealth}
Yiqiang Chen, Xin Qin, Jindong Wang, Chaohui Yu, and Wen Gao.
\newblock Fedhealth: A federated transfer learning framework for wearable
  healthcare.
\newblock {\em IEEE Intelligent Systems}, 2020.

\bibitem{silva2019federated}
Santiago Silva, Boris~A Gutman, Eduardo Romero, Paul~M Thompson, Andre Altmann,
  and Marco Lorenzi.
\newblock Federated learning in distributed medical databases: Meta-analysis of
  large-scale subcortical brain data.
\newblock In {\em 2019 IEEE 16th international symposium on biomedical imaging
  (ISBI 2019)}, pages 270--274. IEEE, 2019.

\bibitem{kim2017federated}
Yejin Kim, Jimeng Sun, Hwanjo Yu, and Xiaoqian Jiang.
\newblock Federated tensor factorization for computational phenotyping.
\newblock In {\em Proceedings of the 23rd ACM SIGKDD International Conference
  on Knowledge Discovery and Data Mining}, pages 887--895, 2017.

\bibitem{liu2019two}
Dianbo Liu, Dmitriy Dligach, and Timothy Miller.
\newblock Two-stage federated phenotyping and patient representation learning.
\newblock In {\em Proceedings of the 18th BioNLP Workshop and Shared Task},
  pages 283--291, 2019.

\bibitem{healthcareIn}
Healthcare in~Europe.
\newblock Federated learning brings ai with privacy to hospitals., 2020.
\newblock Accessed 26 September 2020.
  https://healthcare-in-europe.com/en/home/.

\bibitem{MGB}
Mass~General Brigham.
\newblock Partners healthcare launches federated learning initiative to deliver
  ai at the point of care., 2020.
\newblock Accessed 26 September 2020. https://www.massgeneralbrigham.org/.

\bibitem{mcmahan2017communication}
Brendan McMahan, Eider Moore, Daniel Ramage, Seth Hampson, and Blaise~Aguera
  y~Arcas.
\newblock Communication-efficient learning of deep networks from decentralized
  data.
\newblock In {\em Artificial Intelligence and Statistics}, pages 1273--1282,
  2017.

\bibitem{lin2020real}
S~Lin, G~Yang, and J~Zhang.
\newblock Real-time edge intelligence in the making: A collaborative learning
  framework via federated meta-learning.
\newblock {\em arXiv preprint arXiv:2001.03229}, 2020.

\bibitem{hornik1989multilayer}
Kurt Hornik, Maxwell Stinchcombe, and Halbert White.
\newblock Multilayer feedforward networks are universal approximators.
\newblock {\em Neural networks}, 2(5):359--366, 1989.

\bibitem{rivest1978data}
Ronald~L Rivest, Len Adleman, Michael~L Dertouzos, et~al.
\newblock On data banks and privacy homomorphisms.
\newblock {\em Foundations of secure computation}, 4(11):169--180, 1978.

\bibitem{phong2018privacy}
Le~Trieu Phong, Yoshinori Aono, Takuya Hayashi, Lihua Wang, and Shiho Moriai.
\newblock Privacy-preserving deep learning via additively homomorphic
  encryption.
\newblock {\em IEEE Transactions on Information Forensics and Security},
  13(5):1333--1345, 2018.

\bibitem{FATE}
Webank's AI.
\newblock Federated ai technology enabler., 2020.
\newblock Accessed 26 September 2020. https://www.fedai.org/cn/.

\bibitem{kairouz2019advances}
Peter Kairouz, H~Brendan McMahan, Brendan Avent, Aur{\'e}lien Bellet, Mehdi
  Bennis, Arjun~Nitin Bhagoji, Keith Bonawitz, Zachary Charles, Graham Cormode,
  Rachel Cummings, et~al.
\newblock Advances and open problems in federated learning.
\newblock {\em arXiv preprint arXiv:1912.04977}, 2019.

\bibitem{vincent2010stacked}
Pascal Vincent, Hugo Larochelle, Isabelle Lajoie, Yoshua Bengio, and
  Pierre-Antoine Manzagol.
\newblock Stacked denoising autoencoders: Learning useful representations in a
  deep network with a local denoising criterion.
\newblock {\em Journal of machine learning research}, 11:3371--3408, 2010.

\bibitem{krizhevsky2017imagenet}
Alex Krizhevsky, Ilya Sutskever, and Geoffrey~E Hinton.
\newblock Imagenet classification with deep convolutional neural networks.
\newblock {\em Communications of the ACM}, 60(6):84--90, 2017.

\bibitem{chawla2002smote}
Nitesh~V Chawla, Kevin~W Bowyer, Lawrence~O Hall, and W~Philip Kegelmeyer.
\newblock Smote: synthetic minority over-sampling technique.
\newblock {\em Journal of artificial intelligence research}, 16:321--357, 2002.

\bibitem{elreedy2019comprehensive}
Dina Elreedy and Amir~F Atiya.
\newblock A comprehensive analysis of synthetic minority oversampling technique
  (smote) for handling class imbalance.
\newblock {\em Information Sciences}, 505:32--64, 2019.

\bibitem{yosinski2014transferable}
Jason Yosinski, Jeff Clune, Yoshua Bengio, and Hod Lipson.
\newblock How transferable are features in deep neural networks?
\newblock In {\em Advances in neural information processing systems}, pages
  3320--3328, 2014.

\bibitem{rebuffi2017icarl}
Sylvestre-Alvise Rebuffi, Alexander Kolesnikov, Georg Sperl, and Christoph~H
  Lampert.
\newblock icarl: Incremental classifier and representation learning.
\newblock In {\em Proceedings of the IEEE conference on Computer Vision and
  Pattern Recognition}, pages 2001--2010, 2017.

\bibitem{jung2019unified}
Wonsik Jung, Ahmad~Wisnu Mulyadi, and Heung-Il Suk.
\newblock Unified modeling of imputation, forecasting, and prediction for ad
  progression.
\newblock In {\em International Conference on Medical Image Computing and
  Computer-Assisted Intervention}, pages 168--176. Springer, 2019.

\bibitem{lian2019end}
Chunfeng Lian, Mingxia Liu, Li~Wang, and Dinggang Shen.
\newblock End-to-end dementia status prediction from brain mri using multi-task
  weakly-supervised attention network.
\newblock In {\em International Conference on Medical Image Computing and
  Computer-Assisted Intervention}, pages 158--167. Springer, 2019.

\bibitem{vavoulas2016mobiact}
George Vavoulas, Charikleia Chatzaki, Thodoris Malliotakis, Matthew Pediaditis,
  and Manolis Tsiknakis.
\newblock The mobiact dataset: Recognition of activities of daily living using
  smartphones.
\newblock In {\em ICT4AgeingWell}, pages 143--151, 2016.

\bibitem{grood1983a}
Edward~S Grood and W~J Suntay.
\newblock A joint coordinate system for the clinical description of
  three-dimensional motions: Application to the knee.
\newblock {\em Journal of Biomechanical Engineering-transactions of The Asme},
  105(2):136--144, 1983.

\bibitem{ward2016towards}
Jamie~A Ward, Gerald Pirkl, Peter Hevesi, and Paul Lukowicz.
\newblock Towards recognising collaborative activities using multiple on-body
  sensors.
\newblock In {\em Proceedings of the 2016 ACM International Joint Conference on
  Pervasive and Ubiquitous Computing: Adjunct}, pages 221--224, 2016.

\bibitem{wang2019deep}
Jindong Wang, Yiqiang Chen, Shuji Hao, Xiaohui Peng, and Lisha Hu.
\newblock Deep learning for sensor-based activity recognition: A survey.
\newblock {\em Pattern Recognition Letters}, 119:3--11, 2019.

\bibitem{shi2009mobile}
Guangyi Shi, Cheung~Shing Chan, Wen~Jung Li, Kwok-Sui Leung, Yuexian Zou, and
  Yufeng Jin.
\newblock Mobile human airbag system for fall protection using mems sensors and
  embedded svm classifier.
\newblock {\em IEEE Sensors Journal}, 9(5):495--503, 2009.

\bibitem{he2016smart}
Jian He, Chen Hu, and Xiaoyi Wang.
\newblock A smart device enabled system for autonomous fall detection and
  alert.
\newblock {\em International Journal of Distributed Sensor Networks},
  12(2):2308183, 2016.

\bibitem{yuan2014power}
Jian Yuan, Kok~Kiong Tan, Tong~Heng Lee, and Gerald Choon~Huat Koh.
\newblock Power-efficient interrupt-driven algorithms for fall detection and
  classification of activities of daily living.
\newblock {\em IEEE Sensors Journal}, 15(3):1377--1387, 2014.

\bibitem{he2019low}
Jian He, Zihao Zhang, Xiaoyi Wang, and Shengqi Yang.
\newblock A low power fall sensing technology based on fd-cnn.
\newblock {\em IEEE Sensors Journal}, 19(13):5110--5118, 2019.

\bibitem{chen2017deep}
Min Chen, Xiaobo Shi, Yin Zhang, Di~Wu, and Mohsen Guizani.
\newblock Deep features learning for medical image analysis with convolutional
  autoencoder neural network.
\newblock {\em IEEE Transactions on Big Data}, 2017.

\bibitem{he2020group}
Chaoyang He, Salman Avestimehr, and Murali Annavaram.
\newblock Group knowledge transfer: Collaborative training of large cnns on the
  edge.
\newblock {\em arXiv preprint arXiv:2007.14513}, 2020.

\bibitem{cook2018using}
Diane~J Cook, Glen Duncan, Gina Sprint, and Roschelle~L Fritz.
\newblock Using smart city technology to make healthcare smarter.
\newblock {\em Proceedings of the IEEE}, 106(4):708--722, 2018.

\bibitem{jeon2017automatic}
Hyoseon Jeon, Woongwoo Lee, Hyeyoung Park, Hong Lee, Sang Kim, Han Kim,
  Beomseok Jeon, and Kwang Park.
\newblock Automatic classification of tremor severity in parkinson’s disease
  using a wearable device.
\newblock {\em Sensors}, 17(9):2067, 2017.

\bibitem{menshawy2015automatic}
Mohamed~EL Menshawy, Abdelghani Benharref, and Mohamed Serhani.
\newblock An automatic mobile-health based approach for eeg epileptic seizures
  detection.
\newblock {\em Expert systems with applications}, 42(20):7157--7174, 2015.

\bibitem{chikhaoui2018cnn}
Belkacem Chikhaoui, Frank Gouineau, and Martin Sotir.
\newblock A cnn based transfer learning model for automatic activity
  recognition from accelerometer sensors.
\newblock In {\em International Conference on Machine Learning and Data Mining
  in Pattern Recognition}, pages 302--315. Springer, 2018.

\bibitem{muhammed2018ubehealth}
Thaha Muhammed, Rashid Mehmood, Aiiad Albeshri, and Iyad Katib.
\newblock Ubehealth: a personalized ubiquitous cloud and edge-enabled networked
  healthcare system for smart cities.
\newblock {\em IEEE Access}, 6:32258--32285, 2018.

\bibitem{sametinger2015security}
Johannes Sametinger, Jerzy~W Rozenblit, Roman~L Lysecky, and Peter Ott.
\newblock Security challenges for medical devices.
\newblock {\em Commun. ACM}, 58(4):74--82, 2015.

\bibitem{zhang2019collaborative}
Xingzhou Zhang, Mu~Qiao, Liangkai Liu, Yunfei Xu, and Weisong Shi.
\newblock Collaborative cloud-edge computation for personalized driving
  behavior modeling.
\newblock In {\em Proceedings of the 4th ACM/IEEE Symposium on Edge Computing},
  pages 209--221, 2019.

\bibitem{konevcny2016federated}
Jakub Kone{\v{c}}n{\`y}, H~Brendan McMahan, Felix~X Yu, Peter Richt{\'a}rik,
  Ananda~Theertha Suresh, and Dave Bacon.
\newblock Federated learning: Strategies for improving communication
  efficiency.
\newblock {\em arXiv preprint arXiv:1610.05492}, 2016.

\bibitem{yang2019federated}
Qiang Yang, Yang Liu, Tianjian Chen, and Yongxin Tong.
\newblock Federated machine learning: Concept and applications.
\newblock {\em ACM Transactions on Intelligent Systems and Technology (TIST)},
  10(2):12, 2019.

\bibitem{smith2017federated}
Virginia Smith, Chao-Kai Chiang, Maziar Sanjabi, and Ameet~S Talwalkar.
\newblock Federated multi-task learning.
\newblock In {\em Advances in Neural Information Processing Systems}, pages
  4424--4434, 2017.

\bibitem{lin2017deep}
Yujun Lin, Song Han, Huizi Mao, Yu~Wang, and William~J Dally.
\newblock Deep gradient compression: Reducing the communication bandwidth for
  distributed training.
\newblock {\em arXiv preprint arXiv:1712.01887}, 2017.

\bibitem{chen2018federated}
Fei Chen, Mi~Luo, Zhenhua Dong, Zhenguo Li, and Xiuqiang He.
\newblock Federated meta-learning with fast convergence and efficient
  communication.
\newblock {\em arXiv preprint arXiv:1802.07876}, 2018.

\end{thebibliography}

\ifCLASSOPTIONcaptionsoff
  \newpage
\fi

\end{document}